\documentclass{WileyMSP-template}
\pdfoutput=1 
\usepackage[version=1]{mhchem}
\usepackage[english]{babel}
\usepackage{hyperref}

\begin{document}

\pagestyle{fancy}

\title{Rolled-up Epsilon-near-zero Waveguide reservoir for long-range qubit entanglement}

\maketitle


\author{Ibrahim Issah}
\author{Mohsin Habib}
\author{Humeyra Caglayan*}



\begin{affiliations}
I. Issah, M. Habib, Prof. H. Caglayan\\
Faculty of Engineering and Natural Science, Photonics, Tampere University, 33720 Tampere, Finland.\\
E-mail: humeyra.caglayan@tuni.fi

\end{affiliations}


\keywords{Rolled up metamaterials, ENZ mode, decay rate channels, concurrence, entanglement}

\begin{abstract}
Preservation of the entangled state of a quantum system is relevant in quantum applications. However, the preservation of entangled states is constrained due to the energy dissipation of the quantum system arising from the environment. As a result, the design of the environment seen by quantum bits is relevant due to its relation to the final state of the quantum system. This work presents the concurrence measure of entanglement between two qubits coupled to a rolled-up epsilon-near-zero (ENZ) waveguide reservoir consisting of an alternating layer of metal and dielectric. Our numerical calculations demonstrate that the proposed rolled-up ENZ waveguide reservoir can preserve the entanglement of two qubits at the cutoff wavelength of the reservoir via enhanced energy transfer. This proposed rolled-up ENZ waveguide can serve as a unique reservoir for various quantum technologies such as quantum communication, quantum information processing, and single-photon generation. As a proof of concept, we also demonstrate that this novel structure can be fabricated using cost-effective self-rolling techniques. 
\end{abstract}

\section{Introduction}

Metamaterials are defined as artificially engineered structures with different material properties as compared to naturally existing materials \cite{Liu2011, Nader1}. From their inception, these materials have played enormous roles in the manipulation of electromagnetic fields in many disciplines \cite{Kadic2019, Shalaev2005}. The unique properties of these materials have been identified to serve as means to enhance dipole-dipole interactions, energy harvesting, and long-range interactions of quantum emitters (i.e. quantum dots and diamonds (NV defect centers)) embedded within their waveguide-like meta-structures \cite{Soukoulis2011, Sokhoyan2013, Ren2015, Ding2018}. These physical systems ﬁnd use in many quantum technologies such as quantum communication, quantum information processing \cite{Jha2018}, and single-photon generation \cite{Biehs2016, Loudon}. 

Other unique properties of metamaterials are related to the high enhancement of quantum emitter's response coupled with such a lossy medium. These responses of emitters coupled to such an environment can be described by the dyadic Green’s function which is related to the local density of states formulations. This further leads to enhanced Purcell effects independent of the emitter position along and within the waveguide-like meta-structures. Experimental verifications of such material have been realized for a rectangular ENZ waveguide using cathodoluminescence measurement techniques \cite{Vesseur2013}.  Fleury et al. \cite{Fleury2013}, predominately explored that the flexibility of dipole positions in ENZ waveguide channels at the cutoff wavelength is not the only relevance of these channels but could also boost Dicke superradiance effects which leads to a high collective coherent emission of the quantum emitters. 
 
Plasmonic waveguide channels have also been identified to support extraordinary optical transmission when excited and have been implemented in the subwavelength regime to mediate long-range interactions of quantum emitters \cite{Gonzalez-Tudela2011, DAmico2019}. Recently, Li \textit{et al.} \cite{Li2019} presented a comparative study of ENZ and plasmonic waveguide channels used to enhance efficient long-range resonance energy transfer and inter-emitter entanglement. 
Although plasmonic waveguide types such as V-shaped grooves and cylindrical nanorods have been identified to outperform the sub-wavelength distance limitations of quantum emitters cooperative emission in a homogeneous medium, quantum emitters entangled states mediated by these waveguides suffer from practical applications due to their dependence on the spatial position of emitters \cite{Gonzalez-Tudela2011}. As a result, different techniques have been implemented by many authors to overcome these challenges. 

It is also interesting to note that due to the inherently short-range nature of the dipole-dipole interactions of quantum emitters in a homogeneous medium, it is relevant to envisage different reservoirs that can be used to enhance the cooperative effects of quantum emitters \cite{scully_zubairy_1997}. ENZ waveguide metamaterials have been identified to help in this perspective by exciting modes with zero refractive index which makes it possible to enhance long-range interactions of quantum emitters as well as strong entanglement at farther distances. These metamaterial channels, in particular, exhibit a uniform field amplitude along the waveguide channel which is independent of the axial dipole position due to the integrally large phase velocity \cite{Islam2016, Liu2011, Yildiz2020}. 

These realizations of ENZ waveguide channels have intrigued much interest in the study of decay rate enhancement and cooperative emission of quantum emitters mediated by these channels \cite{Sokhoyan2013}. It is thereby pertinent to practically realize such ENZ waveguide channels with minimal constraints that can be used to enhance resonance energy transfer and long-range entanglement between two-level quantum fluorescence atoms (qubits) \cite{Martin-Cano2011}. However, the difficulty to integrate quantum emitters in planar ENZ waveguide materials in the nanoscale regime has inhibited their practical applications \cite{Li2016}. To fulfill these ENZ features, the integration requires a controllable and feasible 3D fabrication process. The latter, while extremely pertinent from a fundamental perspective, poses limitations due to difficulties in sample fabrication, which may result in reduced ENZ response or create a non-accessible medium for integration and excitation of emitters.

Therefore, in this study, self-assembled three-dimensional rolled-up ENZ waveguides will be our choice to overcome these deficiencies. We designed, fabricated, and numerically simulated alternating layers of metal and dielectric rolled-up ENZ waveguide to serve as an environment to mediate the cooperative emission of emitters embedded within it. We anticipate that the proposed rolled-up ENZ waveguide will enhance long-range dipole-dipole interactions and the preservation of entangled states due to its exotic properties to enhance super coupling within the cutoff region. To study these properties, we implemented the concept of rigorous dyadic Green's function relative to macroscopic quantum electrodynamics (QED) techniques to describe the response of a single fluorescence quantum emitter coupled to the rolled-up ENZ waveguide reservoir. Also, we used the quantum master equation to numerically calculate the transient and steady-state entanglement between two-level atoms mediated by rolled-up ENZ waveguide using Wootters concurrence formalism. It is worth noting that these theoretical calculations are implemented to understand the photonic properties of the ENZ waveguide. The quantum master equation is used to determine the entanglement properties of a single emitter coupled to the central part of the rolled-up ENZ waveguide. Also, the dyadic Green's function is implemented to attain the key coupling parameters of an emitter coupled to the rolled-up ENZ waveguide used in the quantum master equation.

\section{ENZ waveguide modes}
Ostensibly, long-distance entanglement between two quantum bits (qubits) is known to be mediated by photons. However, the recent emergence of the application of surface plasmons generation in different resonators and plasmonic waveguides has attracted researchers to delve into plasmon mediated entanglement between qubits in the nanoscale regime \cite{Fleury2013, Gonzalez-Tudela2011}. This technique to confine optical fields in the subwavelength regime is fundamental in the application of surface plasmons in quantum optics. However, the sinusoidal phase change variations in propagating surface plasmon (SPP) mode of a waveguide channel limit the free distribution of quantum emitters in its corresponding environment \cite{Martin-Cano2010}. It is thereby pertinent to employ alternative means to examine other plasmonic channels with near-zero index and the ability to enhance entanglement between two qubits by coupling with the ENZ waveguide mode.

Here, we investigate the fundamental TE$_{11}$ mode of a rolled-up ENZ structure at the cutoff wavelength where there is minimal phase variation between two quantum emitters. Before we examine the photonic properties of a traditional cylindrical hollow waveguide.

\subsection{Cylindrical hollow waveguide}
To serve as a guide to determine the fundamental mode of the rolled-up ENZ waveguide composed of an alternating layer of metal and dielectric, we numerically implement the analytical equation for a homogeneous circular waveguide with an air core \cite{Snyderbook}. The dispersion relation of the rolled-up ENZ waveguides is initially calculated to identify the cutoff wavelength, where the propagation constant $k = 0$. At this wavelength, the electromagnetic waves can be squeezed or tunneled through a waveguide to exhibit a similar response as ENZ materials. This phenomenon relative to circular waveguides has been demonstrated theoretically by Pan \textit{et al.} \cite{Pan2009} to exhibit unique electromagnetic tunneling which is independent of the waveguide length.

\begin{figure}[!htb]
\includegraphics[width=\linewidth]{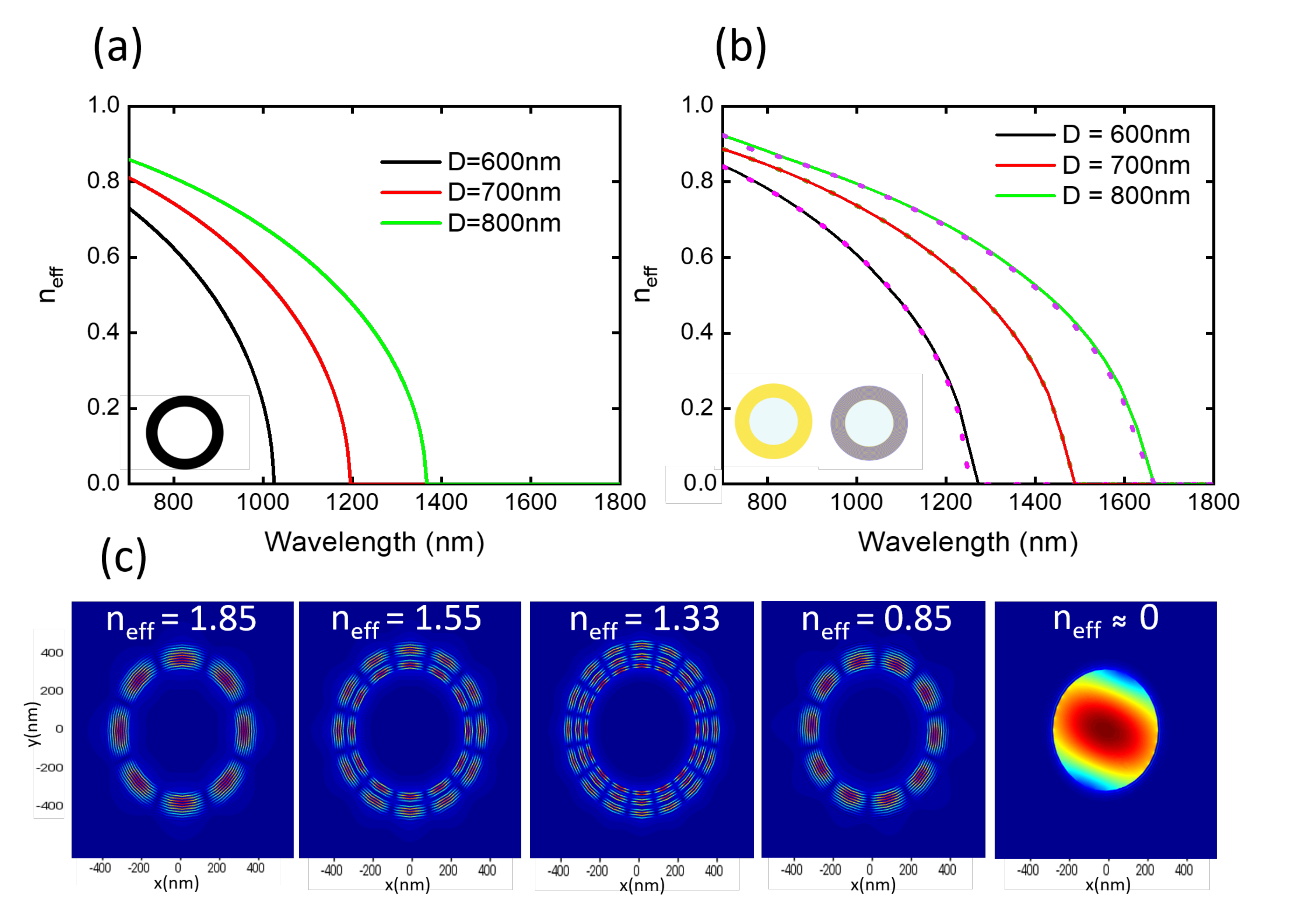}
\caption{\label{fig:fig1ab} (a) The dispersion relation of a cylindrical hollow waveguide with different core diameters (D). (b) The dispersion relation of the rolled-up ENZ waveguide (solid lines). The dotted lines in the same figure shows similar dispersion calculated for the cylindrical hollow metallic waveguide. The cutoff wavelength varies as a function of the core-diameter (D) in both figures (top panel). (c) the different mode profiles excited in the rolled-up ENZ waveguide (D = 700 nm) for different effective indices (\textit{n$_{eff}$}) at the cutoff wavelength ($\lambda = 1450$ nm).}
\end{figure}

From the Helmholtz eigenvalue equation, the cutoff wavelength as a function of the effective index of a circular hollow waveguide can be expressed analytically as
\begin{equation}
n_{e f f}=\sqrt{1-\left(\frac{u_{n m}}{\pi}\right)^{2}\left(\frac{\lambda}{2 \rho}\right)^{2}},
\label{eqn1}
\end{equation} 
where \textit{n$_{eff}$} is the effective index as a function of dielectric core diameter, $u_{n m}$ is the root of the Bessel function, $\rho$ is the radius of the circular core and $\lambda$ is the wavelength. Since our mode of interest is the fundamental TE$_{11}$ mode, the corresponding root of the Bessel function selected is $3.832$. The analytical calculation of the dispersion relation of a cylindrical dielectric waveguide is shown in Fig. \ref{fig:fig1ab} (a). This served as a benchmark to determine the cutoff frequency of the rolled-up ENZ waveguide presented in Figure \ref{fig:fig1ab} (b) for different diameters (D).

For different core diameters of both the cylindrical and ENZ hollow waveguides, we obtain a redshift of the cutoff wavelength. This shows the dependence of the cutoff wavelength on the material dimensions. The dispersion of the cylindrical hollow metallic (i.e. gold (Au)) waveguide (dotted lines) is superimposed in Fig. \ref{fig:fig1ab} (b) shows a similar dispersion relation as the rolled-up ENZ waveguide. Figure \ref{fig:fig1ab} (c) illustrates the unique and complex modes of the rolled-up ENZ waveguide at different effective indices ($n_{eff}$). The complex mode profiles with effective cladding index (\textit{n$_{eff}$} $\neq$ 0) are mostly confined in the metal-dielectric cladding region due to the excitation of plasmon modes. However, the fundamental mode of the waveguide with \textit{n$_{eff}$} $\simeq$ 0 confined in the core possesses a similar dispersion relation with the fundamental modes of the plasmonic hollow waveguide. This waveguide fundamental mode with an effective index near zero introduces novel ways of controlling optical field propagation and enhancement due to the infinite phase velocity of the tunneled electromagnetic wave.

\subsection{Rolled-up ENZ waveguide}
In this section, as a proof of design concept, we described how the proposed structure can be fabricated. 

The first rolled-up tubes were fabricated by Prinz \textit{et al.} using strained InAs/GaAs bilayer with lattice mismatch \cite{PRINZ2000828}. The layers start to roll as the sacrificial layer beneath them is released by an etchant. However, semiconductor-based rolled-up tubes are not suitable for our purpose, due to the high refractive indices of material as compared to near-zero refractive index material required for this study. In this study, we adopted a similar strained induced self-rolling mechanism to obtain a three-dimensional rolled-up ENZ waveguide of Au and SiO$_{2}$ on a silicon (Si) substrate using germanium (Ge) as a sacrificial layer (see Appendix for details). Figure \ref{fig:fig1schem} (a) illustrates the schematics of the proposed design. Figure \ref{fig:fig1schem} (b) depicts the scanning electron microscope (SEM) image of the fabricated rolled-up ENZ waveguide. The structure has a core diameter (D) of 700 nm, comprising of 12 alternating bi-layers and 10 nm of Au and 5 nm of glass silica (SiO$_{2}$) thick, and length (L) of 25 $\mu$m.

The benefit of the rolled-up ENZ waveguide over other plasmonic waveguide channels is that it offers different emitters integration techniques into the core of the waveguide. For example, injection techniques of colloidal nanoemitters using microsyringe technique. In such a case, the driving of nanoemitters is mediated by capillary forces \cite{Strelow2012}. Another method of integrating quantum emitters to the core of the rolled-up ENZ waveguide is to deposit emitters on the planar bilayer before initiating the rolling process.
 
We now move to the numerical calculations based on the parameters of the fabricated structure. However, due to the existence of the non-radiative and propagation losses, the 25 $\mu$m length of the fabricated waveguide is approximated to be 3 $\mu$m in the simulation. Moreover, the 3 $\mu$m  was identified to suffice for determining the coupling parameters of an emitter positioned at the central part of the rolled-up ENZ waveguide. It is also interesting to note that the numerical calculation suggests that for a dipole placed at the central part of the waveguide, its decay rate vanishes beyond 3 $\mu$m as we shall see in Fig. \ref{fig:fig5}.

The core diameter of the rolled-up structure is filled with a material permittivity of one. Material dispersion of Au from Johnson and Christy \cite{Johnson1972} material dispersion data and Palik \cite{palik1997handbook} data for the SiO$_{2}$ layers were used in the modeling. Figure \ref{fig:fig1schem} (c) also shows the volumetric display of the fundamental mode of the rolled-up ENZ waveguide at the cutoff wavelength $\lambda \simeq 1450$ nm. The embedded dipoles represent the quantum emitters mediated by the ENZ mode of the rolled-up ENZ waveguide and the vector surface plot shows the field distribution of the fundamental TE$_{11}$ mode.

\begin{figure}[!htb]
\includegraphics[width=\linewidth]{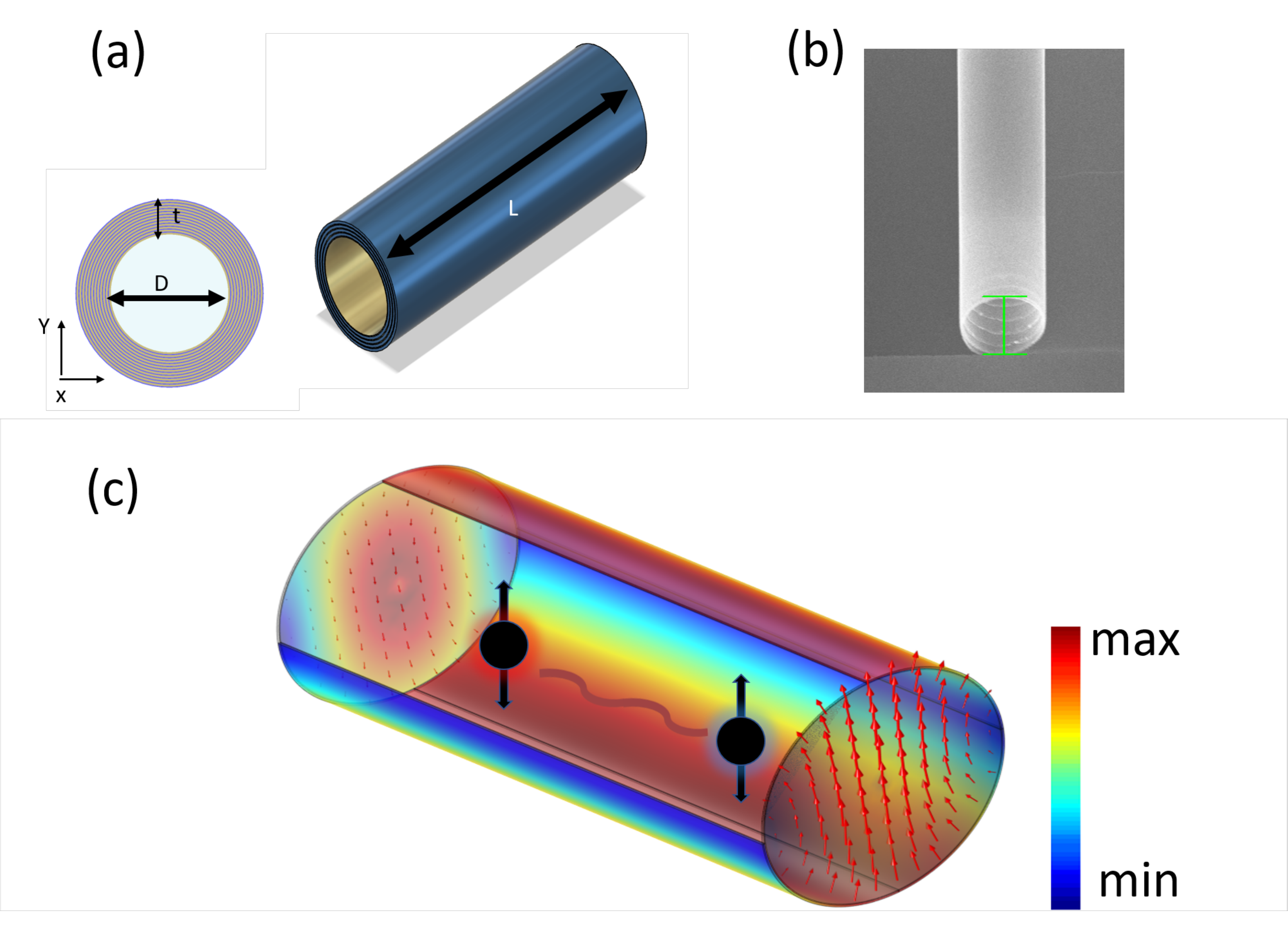}
\caption{\label{fig:fig1schem} (a) Schematic of the rolled-up ENZ waveguide (b) SEM images of the fabricated waveguide. The green line corresponds to a diameter of 700 nm and the length of the waveguide is 25 $\mu$m. (c) the volumetric display of the normalized fundamental TE$_{11}$ mode propagation in the rolled-up ENZ waveguide. The embedded dipoles (qubits) are denoted by the black spheres with arrows. The interaction between the dipoles mediated by the ENZ mode is represented by the curvy lines. The surface vector plot shows the field distribution of the fundamental mode.}
\end{figure}

\section{Decay rate enhancement of the rolled-up waveguide}
As stated initially, advances in plasmonic materials (i.e. V-shaped grooves, cylindrical nanorods) and  ENZ plasmonic metamaterials (i.e. plasmonic planar waveguide) have availed the opportunity to enhance superradiant effects from a collective quantum emitter which outperforms the weak interaction of emitters in a homogeneous medium \cite{Fleury2013}. The superradiance effect known as the collective effect of quantum emitters arranged close to each other was proposed by Dicke \cite{Dicke1954} to show the relationship between the radiation intensity of quantum emitters and the number of quantum sources. This radiant effect is linked to the exotic properties of the reservoir to which the quantum emitters are coupled to.

At the cutoff wavelength region of the rolled-up ENZ waveguide channel, we foresee a high local density of state (LDOS) of the quantum emitter embedded in the waveguide structure. This enhancement at the cutoff wavelength is insensitive to the emitter axial position and thereby exhibits the aforementioned inherent flexibility of emitters position in ENZ metamaterials \cite{Ozgun2016, Ozgun2018, Li2019}.

\begin{figure}[!htb]
\includegraphics[width=\linewidth]{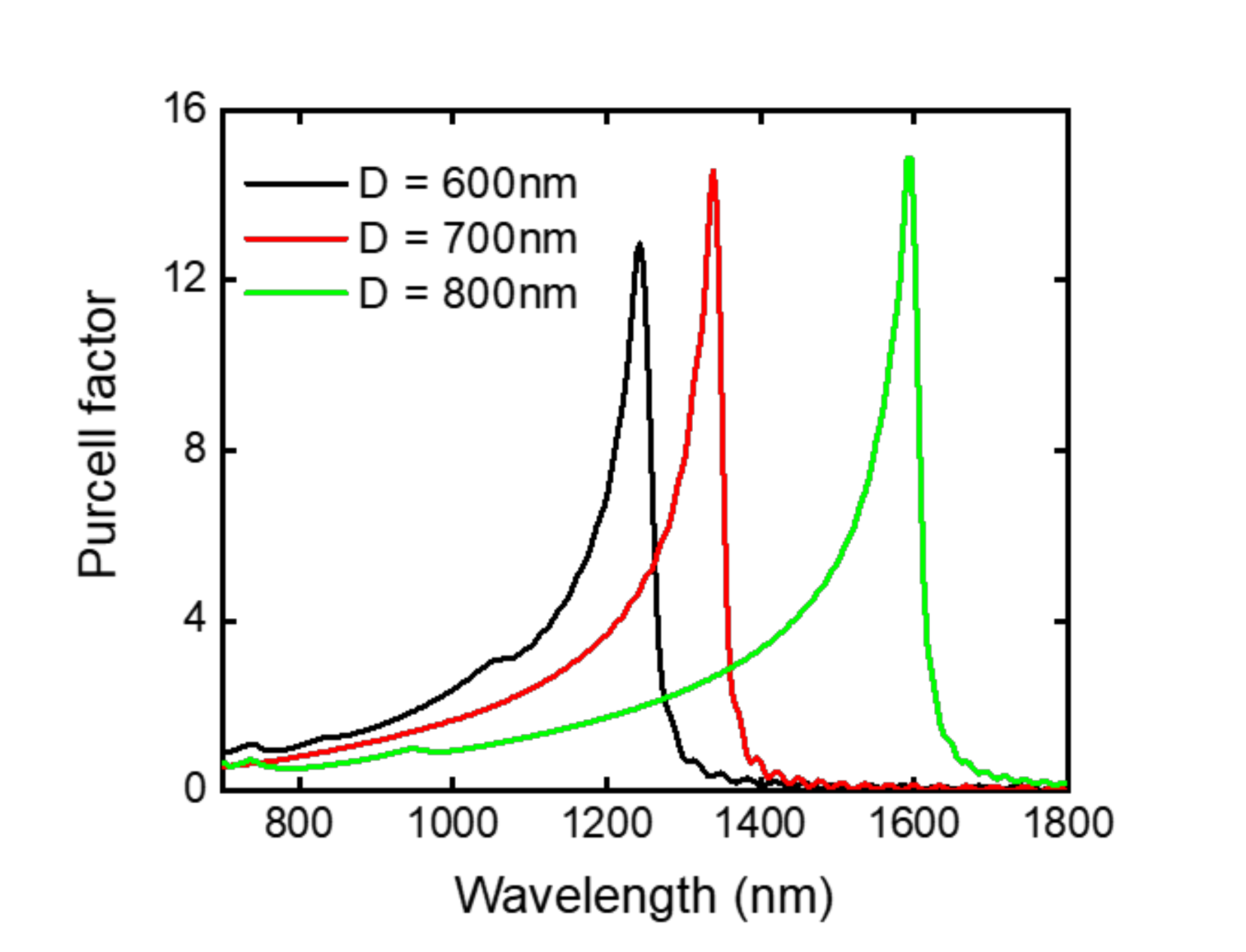}
\caption{\label{fig:fig1a}Decay rate enhancement for different core diameters of the rolled-up ENZ waveguide.}
\end{figure}

To verify the high LDOS at the cutoff wavelength, we computed the Purcell factor of the proposed rolled-up ENZ waveguide channels as a function of different waveguide core diameters. Figure \ref{fig:fig1a} presents the corresponding Purcell factor calculation. It can be seen that the spectral resonance response of a single quantum emitter coupled to a rolled-up ENZ waveguide is dependent on the core diameter. As the diameter increases the spectral density response redshifts to a higher wavelength similar to the dispersion relation in Fig. \ref{fig:fig1ab} (b).  For the waveguide channel with a core diameter of $700$ nm, the peak enhancement is around the cutoff wavelength (i.e. $\lambda $ $\simeq 1450$ nm).
 
Above the cutoff wavelength, the decay rate is predominantly quenched by the environment as there is a very low density of states due to the quenched propagating modes in the waveguide. However, below the cutoff wavelength, there is monotonic build-up in the decay rate reaching a maximum at the cutoff wavelength. Close to the cutoff wavelength, the decay rate is not only enhanced but remains uniform along the waveguide channel. This uniformity is due to the non-resonant mode distribution at the cutoff wavelength \cite{Vesseur2013}. 

\section{Entanglement theory of qubits coupled with an ENZ waveguide}
To determine the long-range and duration of qubits interactions mediated by the rolled-up ENZ waveguide, the dyadic response of the ENZ waveguide at the cutoff wavelength is calculated and compared with two other wavelengths. We also introduce the theory of steady-state and transient entanglement of qubits using the concurrence metric using Wootters formalism \cite{Wootters1998} as a basis for the measure of entanglement. 

\subsection{Coupling parameters} 
We describe here the evolution of a quantum system coupled to the rolled-up ENZ waveguide by introducing the dyadic Green’s function to determine the coupling parameters \cite{Dzsotjan2011, Dung2002}. The coupling parameters help in determining the behavior of quantum two-level emitters (qubits) placed inside a lossy medium using the quantum master equation \cite{Dzsotjan2011, Dung1998, DAmico2019}. Although the dyadic Green’s function is a classical quantity, yet it has been used quite extensively to study the spontaneous decay and dipole-dipole interactions of emitters coupled to either lossy or lossless materials \cite{Li2016, Ding2018}.

The electric field dyadic Green’s function is described by a 3-by-3 matrix with each column representing the orientation of the dipole source. Using the Finite-Difference Time-Domain (FDTD) Lumerical software, we attained the Green’s function of a quantum emitter embedded in the central part of the rolled-up ENZ waveguide which result to a dissipative decay rate ($\gamma_{i j}$) and coherent dipole-dipole coupling term ($g_{i j}$) expressed as
\begin{equation}
\gamma_{i j}=\left(2 \omega_{0}^{2} / \varepsilon_{0} h c^{2}\right) \operatorname{Im}\left[\boldsymbol{\mu}_{i}^{*} \cdot \mathbf{G}\left(\mathbf{r}_{i}, \mathbf{r}_{j}, \omega_{0}\right) \cdot \boldsymbol{\mu}_{j}\right]
\label{eqn2}
\end{equation}
\begin{equation}
\left.g_{i j}=\left(\omega_{0}^{2} / \varepsilon_{0}  h c^{2}\right) \operatorname{Re} \mid \boldsymbol{\mu}_{i}^{*} \cdot \mathbf{G}\left(\mathbf{r}_{i}, \mathbf{r}_{j}, \omega_{0}\right) \cdot \boldsymbol{\mu}_{j}\right],
\label{eqn3}
\end{equation}

where $\mu$ is the quantum emitter dipole moment, $h$ is the Planck's constant, $\epsilon_{0}$ is the free space permittivity, $c$ is the speed of light, and $G$ is the dyadic Green's function. $\gamma_{i i}$ represents the dissipative decay rate of the qubit within a reservoir including the rolled-up ENZ waveguide. $\gamma_{i j}$ represents the decay rate of an emitter at position $r_{i}$ that induces dipole interactions, mediated by an ENZ mode, with a corresponding emitter at $r_{j}$. The transition frequency shift $\omega_{0}$ induced by the dipole-dipole interaction of two qubits embedded within the rolled-up ENZ waveguide is represented by $g_{i j}$. Also, photonic Lamb shift $g_{i i}$ coined as the self-interaction of a quantum emitter placed at the central part of the rolled-up ENZ waveguide is assumed to be negligible. Li \textit{et. al} \cite{Li2019} also emphasized that Lamb shift $g_{ii}$ can be neglected when the distance between an emitter and a reservoir is longer than $10$ nm. 

By considering an emitter placed in the center of a rolled-up ENZ waveguide with symmetry conditions, we also assumed that the total dissipative rate can be expressed as $\gamma_{T}$ = $\gamma_{rad}$ + $\gamma_{Joule}$ + $\gamma_{ENZ}$, based on the emitter decay channels (i.e. free-space radiance ($\gamma_{rad}$),  ohmic losses in metals ($\gamma_{Joule}$), and excitation of the fundamental transverse electric (TE) ENZ mode ($\gamma_{ENZ}$).

Particularly, the free-space radiative decay channel is small as compared with the long-range entanglement mediated by the ENZ mode and thereby neglected. Also, the presence of the ohmic losses is deleterious and can also be neglected since it does not contribute much to the long-range emitters interactions. The inherent losses of the ENZ mode are also accounted for in the dyadic Green's function. The coupling parameters utilized in the quantum master equation, described below in section 4.2, accounts for an emitter decay channel due to its interaction with the electromagnetic field of the rolled-up ENZ waveguide reservoir. However, the existence of other non-radiative channels of the emitter is unavoidable. Typically, quantum dots electron-phonon interaction exhibits non-radiative quenching of emitters photoluminescence \cite{Gangaraj2015}. Due to the exotic properties of ENZ materials highlighted initially, the axial position of the dipole emitter inside the rolled-up ENZ waveguide is not restrained and is identified to have negligible quenching channels.

\subsection{Quantum master equation} 
After identifying the key coupling parameters in the quantum master equation, we further determined the transient and steady-state entanglement of the two qubits inside the rolled-up ENZ waveguide. The master equation is implemented to describe the evolutionary dynamics of the density matrix $\rho$ of the two-qubit system in the locality of a reservoir including the rolled-up ENZ waveguide \cite{Gangaraj2017}.
Assuming weak excitation and implementing the Born Markov and rotating wave approximation \cite{Gangaraj2015}, the master equation can be represented as
\begin{equation}
\frac{\partial \rho}{\partial t}=\frac{1}{i \hbar}[H, \rho]-\frac{1}{2} \sum_{i, j=1}^{2} \gamma_{i j}\left(\rho \sigma_{i}^{\dagger} \sigma_{j}+\sigma_{i}^{\dagger} \sigma_{j} \rho-2 \sigma_{i} \rho \sigma_{j}^{\dagger}\right)
\label{eqn7}
\end{equation}
where the Hamiltonian accounts for the coherent part of the dynamics expressed as 
\begin{equation}
H=\sum_{i} h\left(\omega_{0}+g_{i i}\right) \sigma_{i}^{\dagger} \sigma_{i}+\sum_{i \neq j} \hbar g_{i j} \sigma_{i}^{\dagger} \sigma_{j}.
\label{eqn8}
\end{equation}
From Eqn (\ref{eqn7}) $\rho$ shows the density matrix of the quantum system of two qubits ($i, j)$ with identical transition frequency $\omega_{0}$. $\sigma^{\dagger}$ and $\sigma$ represent the creation and destruction operators applied to the corresponding qubit, respectively. It is interesting to note that the decay rate excitations in the vicinity of a reservoir are higher than the decay rate of the emitter excited state. Therefore, the Born and Markov approximations are considered valid for the quantum master equation.

\begin{figure}[!htb]
\centering
\includegraphics[width=.5\linewidth]{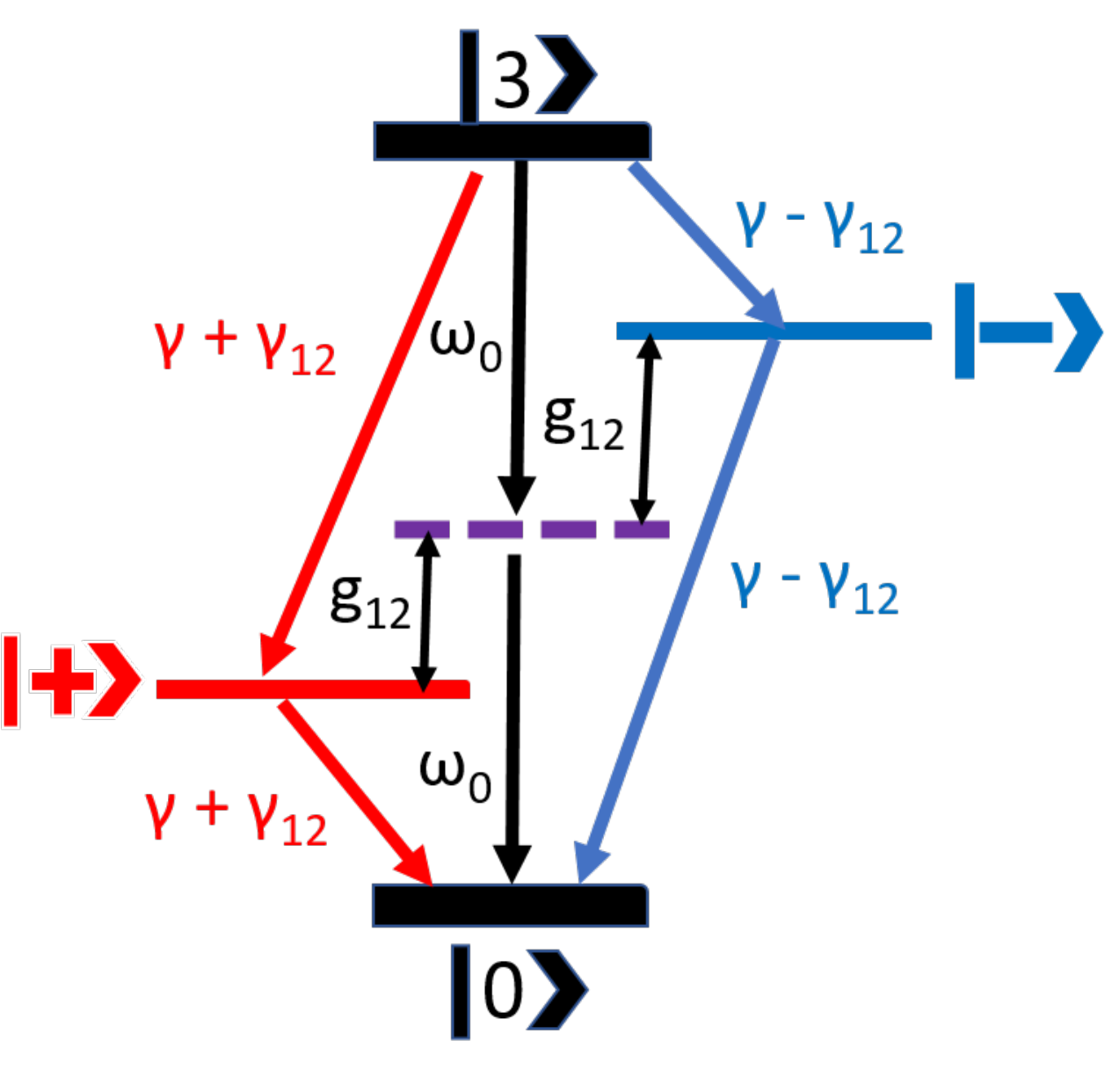}
\caption{\label{fig:figentstate} Collective states of two identical atoms with arrows indicating the possible one-photon transition. The energies of the symmetric $|+\rangle$ and antisymmetric $|-\rangle$ states are shifted by the dipole–dipole interaction $\pm g_{12}$. The coupling between the two atoms induces the collective superradiant $\gamma + \gamma_{12}$ and subradiant $\gamma -\gamma_{12}$ decay rates. $\omega_{O}$ indicates identical transition frequency of the two emitters.}
\end{figure}

The master equation can then be solved analytically on a convenient basis for the two-qubit system vector space. By considering identical qubits with the same transition frequency placed in the same position, i.e. $\gamma_{11}=\gamma_{22}=\gamma$, the Dicke basis can be expressed as

\begin{equation}
|3\rangle=\left|e_{1}, e_{2}\right\rangle,|0\rangle=\left|g_{1}, g_{2}\right\rangle
\label{eqn9}
\end{equation} 
\begin{equation}
|\pm\rangle=1 / \sqrt{2}\left(\left|e_{1}, g_{2}\right\rangle \pm\left|g_{1}, e_{2}\right\rangle\right),
\label{eqn10}
\end{equation}
where $\left|g_{i}\right\rangle$ and $\left|e_{i}\right\rangle$ are the ground and excited states of the $i$ -th qubit state, respectively. These basis are sufficient enough to represent the dynamics of the quantum system. Fig. \ref{fig:figentstate} illustrates the collective states of two identical emitters coupled to a lossy medium \cite{Ficek2014}.
It can be seen that the dipole-dipole interaction $g_{12}$ does not affect the excited $|3\rangle$ and ground $|0\rangle$ states, but shifts the energies of the symmetric $|+\rangle$ and antisymmetric $|-\rangle$ states. This in turn induces the collective superradiant decay $\gamma + \gamma_{12}$ and subradiant $\gamma -\gamma_{12}$ decay rates.

Using this collective decay formalism and assuming that only one of the emitters is initially excited such that $\rho_{33}(t)=0$ and $\rho_{00}(0)=0$, we can produce an effective entanglement approach. Hence the quantum system can be prepared in an unentangled basis state $\left|e_{1}, g_{2}\right\rangle=1 / \sqrt{2}(|+\rangle+|-\rangle)$ such that the density matrix with non-zero terms are $\rho_{++}(0)=\rho_{+-}(0)=\rho_{-+}(0)=\rho_{--}(0)=1 / 2$ \cite{Gangaraj2015}. Using these initial conditions of the density matrix dynamics, the concurrence can be expressed as 
\begin{equation}
C(t)=\sqrt{\left[\rho_{++}(t)-\rho_{--}(t)\right]^{2}+4 \operatorname{Im}\left[\rho_{+-}(t)\right]^{2}},
\label{eqn11}  
\end{equation}
which takes the form
\begin{equation}
C(t)=\frac{1}{2} \sqrt{\left[e^{-\left(\gamma+\gamma_{12}\right) t}-e^{-\left(\gamma-\gamma_{12}\right) t}\right]^{2}+4 e^{-2 \gamma t} \sin ^{2}\left(2 g_{12} t\right)}.
\label{eqn12}
\end{equation}

In general, to characterize the entanglement between two quantum emitters, the concept of concurrence introduced by Wootters \cite{Wootters1998} is implemented which is defined as 
\begin{equation}
C=\max \left(0, \sqrt{u_{1}}-\sqrt{u_{2}}-\sqrt{u_{3}}-\sqrt{u_{4}}\right),
\label{eqn13}
\end{equation}
where $u_{i}$ represents the eigen values of the matrix $ \rho \tilde{\rho}$ and $\tilde{\rho}=\sigma_{y} \otimes \sigma_{y} \rho^{*} \sigma_{y} \otimes \sigma_{y}
$ is the spin-flip density matrix with $\sigma_{y}$ being the Pauli matrix. The degree of concurrence is determined between 1 (completely entangled state) and 0 (unentangled state).
Equation (\ref{eqn12}) shows the derived transient entanglement between two qubits which is dependent on time. At the time $t=0$, the concurrence $C(0)=0$ as expected, since only one emitter is excited and thereby the initial state is unentangled. At $t>0$, the quantum emitters become entangled as a function of an increase in the concurrence values. 

Notably, a lossless and infinite waveguide with $\gamma = \gamma_{12}$ shows a monotonic growth of concurrence with a maximum value of 0.5 based on the initial conditions of the density matrix presented in Eqn. (\ref{eqn12}). However, this effect is transient and does not last for a long period. Thus, to obtain a steady entangled state, an external pumping is required to compensate for the depopulation of the qubit excited states. This effect of excited state depopulation is a direct consequence of decoherence due to radiative and non-radiative losses. By restricting the external pump to a classical monochromatic source, the electric field $\mathbf{E}_{\alpha}=\mathbf{E}_{0 \alpha} e^{-i \omega_{p} t}+\mathbf{c . c .}$ adds an additional perturbation $(i / h)\left[V,\rho_{s}(t)\right]$ to the master equation where the operator
\begin{equation}
V=-\sum_{i}^{2} \hbar\left(\Omega_{i} e^{-i \Delta_{i} t} \sigma_{i}^{\dagger}+\Omega_{i}^{*} e^{i \Delta_{i} t} \sigma_{i}\right),
\label{eqn14}
\end{equation}
accounts for the interaction between the qubits and the pump source with an effective Rabi frequency $\Omega_{i}=\mu \cdot E_{0 i} / h$. The parameter $\Delta_{i}=\omega_{0}-\omega_{L}$ is the detuning frequency between the pump field and the transition frequency of the qubits. Concerning rolled-up ENZ waveguides, one can mainly excite emitters embedded within the waveguide channel at a specific ENZ spectral resonance wavelength due to its nature which results in a zero (0) detuning parameter ($\Delta_{i}= 0$). The coupling of an electromagnetic field within such an ENZ waveguide channel is around a particular resonance wavelength. The above formulations were used in generating systems of equations to solve the quantum master equation analytically and to determine the concurrence of the qubits as a measure of the entanglement.

\subsection{ Rolled-up ENZ waveguide dipole-dipole coupling and decay rate}
As stated initially, to determine the key parameters (i.e. $\gamma_{12}$, $\pm g_{12}$) to solve the quantum master equation in Eqn. (\ref{eqn7}), we calculate the decay rate and dipole-dipole interaction of an emitter embedded within the rolled-up ENZ waveguide using the dyadic Green's function obtained from the FDTD simulation. We study these ingredients of the quantum master equation at three different wavelengths (i.e. the cutoff wavelength ($\lambda = 1450$ nm) and at two other wavelengths: close ($\lambda = 1300$ nm) and far ($\lambda = 1250$ nm) from the cutoff wavelength). 

\begin{figure}[!htb]
\includegraphics[width=\linewidth]{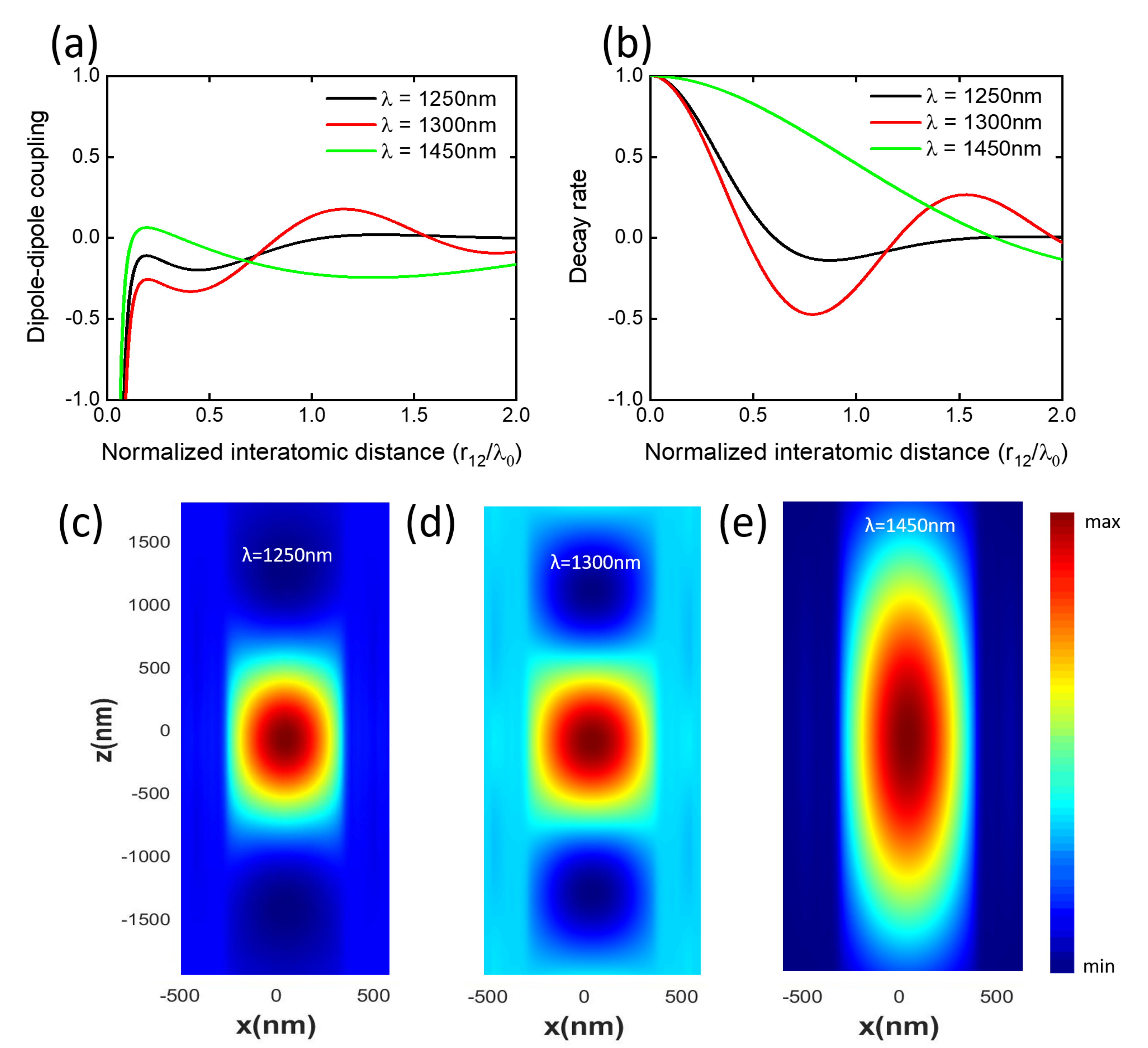}
\caption{\label{fig:fig5}Coupling parameters of the rolled-up ENZ waveguide. (a) the dipole-dipole interaction and (b) the decay rate of emitters coupled to the rolled-up ENZ waveguide at different excitation wavelengths. Energy transfer resonance (ETR) of the rolled-up ENZ waveguide for the different excitation wavelengths: (c) 1250 nm (d) 1300 nm (e) 1450 nm.}
\end{figure}

Figure \ref{fig:fig5} (a) depicts the dipole-dipole interactions of quantum emitters embedded in a rolled-up ENZ waveguide at the three different wavelengths as a function of interatomic distance ($r_{12}/\lambda_{0}$). At the cutoff wavelength, the dipole interactions increase appreciably at short-range and decay exponentially as a function of interatomic distance as compared with the other two wavelengths. At $\lambda = 1250$ nm and $\lambda = 1300$ nm, we identify an oscillatory behavior of the dipole-interactions as compared with the cutoff wavelength. This oscillatory behavior of the two wavelengths relative to the cutoff wavelength is also shown in the spontaneous decay rate of the quantum emitter as shown in Fig. \ref{fig:fig5} (b). Figure \ref{fig:fig5} (c)-(e) presents the energy transfer resonance, which is dependent on the dyadic Green's function, of the qubits inside the rolled-up ENZ waveguide \cite{Ren2016}. This shows that, at the cutoff wavelength, the rolled-up ENZ waveguide provides modes with a longer wavelength to enhance strong entanglements of two qubits placed at farther distances.

At the cutoff wavelength $\lambda_{0} = 1450$ nm, we also see a long-range decay rate as a function of interatomic distance which enhances qubit entanglement. It is evident that the comparison between the normalized decay rate and the dipole-dipole interaction at the cutoff wavelength satisfies the condition of attaining high entanglement performance i.e., $g_{12} \ll \gamma$ and $\gamma=\gamma_{12}$ as shown in Fig. \ref{fig:fig5} (a) and (b). Note, that the emitter's decay time at the cutoff wavelength is normalized by its maximum value of approximately $10^{-5}$ s at $r_{12}/\lambda_{0} = 0$. The maximum value of the decay time is represented by the qubit self-interaction term $\gamma$. We also observed that $r_{12}/\lambda_{0} = 0$ to  $r_{12}/\lambda_{0} = 1$, there is an appreciable high decay ratio $\gamma_{12} / \gamma$ which results in a suppression of the subradiant decay state $|-\rangle$ as compared with the superradiant $|+\rangle$ state (i.e. superradiant $\gamma + \gamma_{12}$ and subradiant $\gamma -\gamma_{12}$ decay rates). Based on the coupling parameters ((Fig. \ref{fig:fig5} (a, b)) and energy transfer resonance (Fig. \ref{fig:fig5} (e)) predicts strong long-range interaction of qubits and high persistence of entangled states in the rolled-up ENZ waveguide at the cutoff wavelength. 

\subsection{Measure of entanglement}
Using the decay rate and the dipole-dipole interactions of the qubits inside the rolled-up ENZ waveguide, we calculated the corresponding concurrence metric measure of entanglement (transient and steady-state) using Eqn. (\ref{eqn12}) and (\ref{eqn13}). This measure of entanglement is relevant to determine the long-range interactions of quantum emitters and their duration.
\subsubsection{Transient Entanglement mediated by rolled-up ENZ waveguide}
Using the cutoff wavelength $\lambda_{0}$ and self-interactions $\gamma$ as a normalization factor for both the interatomic distance $r_{12}$, and evolution time $t$, we calculated the concurrence metric of the quantum emitters as a function of normalized time  $\gamma t$ and interatomic distance $r_{12}/\lambda_{0}$ at different wavelengths using Eqn. (\ref{eqn12}). 

\begin{figure}[!htb]
\includegraphics[width=\linewidth]{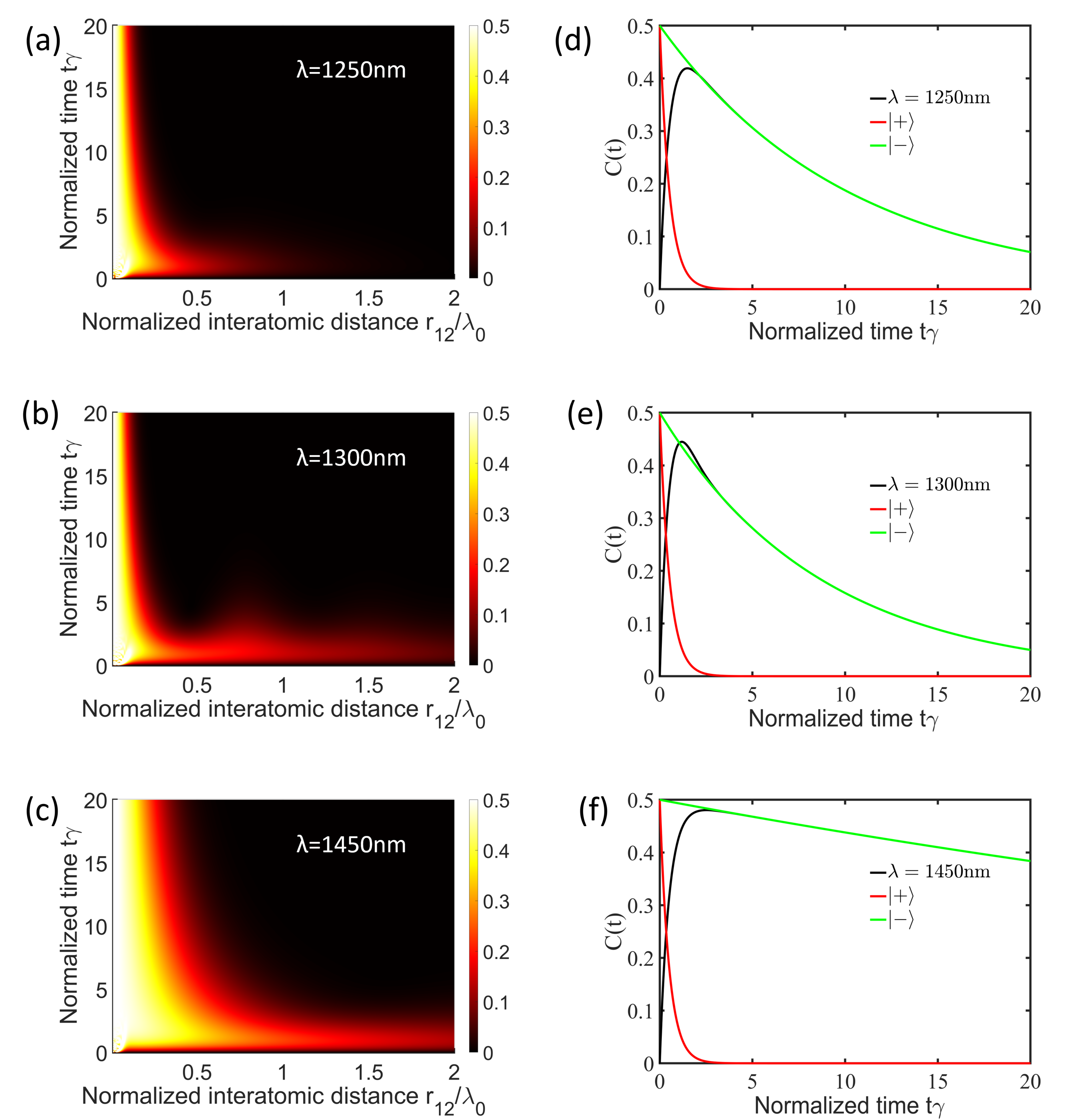}
\caption{\label{fig:fig7}The measure of entanglement between qubits embedded within the rolled-up ENZ waveguide. (a) - (c) the concurrence metric heatmap for different wavelengths: (a) $1250$ nm (b) $1300$ nm and (c) $1450$ nm (ENZ wavelength). (d) - (e) similar concurrence metric plot at normalized interatomic distance ($r_{12}/\lambda_{0} = 0.1$) and their corresponding symmetric $|+\rangle$ and antisymmetric $|-\rangle$ states.}
\end{figure}

Figure \ref{fig:fig7} (a) - (c) illustrates the measure of entangled states using the concurrence metric formalism for the different wavelengths. It is evident that the concurrence metric is higher at the cutoff wavelength as compared to the other two wavelengths. The high concurrence as a function of time and inter-emitter distance is due to the excitation of the ENZ mode of the waveguide which mediates the long-range interactions of the quantum emitters. At $\lambda = 1300$ nm, which is closer to the cutoff wavelength ($\lambda = 1450$ nm), shows a higher entanglement as compared to the concurrence for $\lambda = 1250$ nm. 

The excited mode at $\lambda = 1250$ nm wavelength region has efficient entanglement at short-range but decreases monotonically as a function of inter-emitter distance due to the confined field as shown in Fig. \ref{fig:fig5} (c). However, at the cutoff wavelength, we obtain a large homogeneous electromagnetic field as shown in Fig. \ref{fig:fig5} (e). This results in a large decay rate value and small dipole-dipole interactions and thereby leads to a higher concurrence independent of the emitter position. To appreciate the latter effect, we plotted the concurrence metric at $r_{12}/\lambda_{0} = 0.1$ as a function of normalized time $\gamma t$ for the different wavelengths i.e. $\lambda = 1250$ nm, $\lambda = 1300$ nm, and $\lambda_{0} = 1450$ nm, respectively as shown in Fig. \ref{fig:fig7} (d), (e), and (f). It is clear that at the cutoff wavelength, we have high concurrence which persists for an appreciable time as compared to the other wavelengths. This effect is a consequence of the suppressed subradiant state $|-\rangle$ as compared with the high superradiant $|+\rangle$ decay channels which are superimposed on the concurrence plot in Fig. \ref{fig:fig7} (d), (e), and (f). Clearly, a decrease in the population dynamics of the subradiant state $|-\rangle$ at $\lambda = 1250$ nm and $\lambda = 1300$ nm wavelengths results in a lower persistence of entanglement over time. Comparatively, the dependence of the subradiant $|-\rangle$ population of states is appreciably high at the cutoff wavelength which leads to a high concurrence state. 

\subsubsection{Steady-state entanglement mediated by rolled-up ENZ waveguide }
Up until now, we have justified the relevance of concurrence metric key parameters (i.e. $\gamma_{12}$, $\pm g_{12}$) which can be beneficial to achieve appreciable entanglement useful for quantum information processing and communication. We have also demonstrated the concurrence as a function of normalized time and interatomic distance as well as illustrate the flexibility of emitter axial positions at the cutoff wavelength for the rolled-up ENZ waveguide. However, this entanglement decays with time as shown in Fig. \ref{fig:fig7} (d) - (f). In the case of Fig. \ref{fig:fig7} (f), we can visualize the reduction in concurrence as a function of the gradual monotonic decay of the subradiant $|-\rangle$ population of states as compared with the sharp decay of Fig.\ref{fig:fig7} (d) and (e). This transient effect of entanglement between emitters embedded within a rolled-up ENZ waveguide is due to decoherence which is attributed to radiation losses and depopulation of emitters in the excited states \cite{Li2019}. To attain a steady entangled state $C(t \rightarrow \infty)$, it is relevant to compensate for the depopulation of the excited states by pumping the qubits with an external source. It is also interesting to note that to achieve strong steady-state entanglement between the qubits, the pump strength should not be too large; otherwise, strong interactions between the pump and the qubits as well as the pump and the ENZ channel may occur. This could eventually lead to qubit decoupling and lasing or strong resonances which may affect the decay channel of the emitter embedded within the waveguide channel \cite{Gangaraj2017}. In the current case, we utilized a weak pumping scenario and considered the effect of the pump on the ENZ waveguide negligible. We used pump intensities optimal to compensate for the depopulation of emitters in the excited states to enhance long-range entangled states.

\begin{figure}[!htb]
\includegraphics[width=\linewidth]{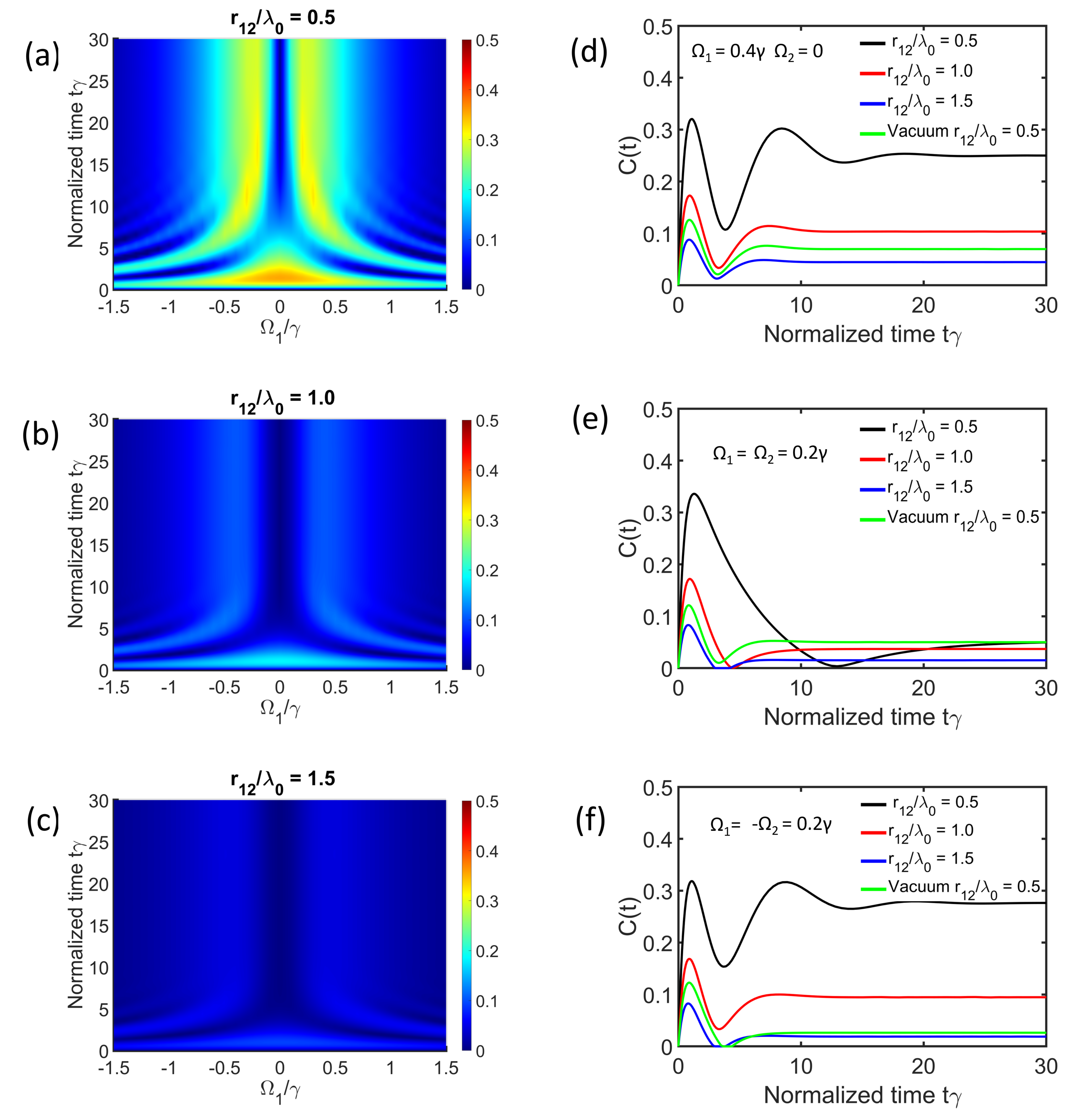}
\caption{\label{fig:fig7a} Persistent time dependent concurrence as a function of an external coherent source. The time dependent concurrence for different Rabi frequencies $\Omega_{1}/\gamma$ of a single pump source ($\Omega_{2}/\gamma$ = 0) for different normalized interatomic distances (a) $r_{12}/\lambda_{0} = 0.5$, (b) $r_{12}/\lambda_{0} = 1.0$, and (c) $r_{12}/\lambda_{0} = 1.5$. Time dependent concurrence between two quantum emitters at different normalized interatomic distances for different external pumping (d) asymmetric ($\Omega_{1} = 0.4\gamma, \Omega_{2}=0$), (e) symmetric ($\Omega_{1}=\Omega_{2} = 0.2\gamma$), and (f) antisymmetric ($\Omega_{1}=-\Omega_{2} = 0.2\gamma$). Similar plot of the homogeneous medium is illustrated by the plot legend Vacuum $r_{12}/\lambda_{0} = 0.5$}
\end{figure}

We also assume a detuning parameter $\Delta_{i}=\omega_{0}-\omega_{\mathrm{p}}=0$ by utilizing an external source with a resonance frequency $\omega_{p}$ similar to the transition frequency $\omega_{0}$ of the quantum emitters embedded in the rolled-up ENZ waveguide. Figure \ref{fig:fig7a} (a) - (c) shows the heatmap of the steady-state concurrence as a function of varied normalized Rabi frequencies $\Omega_{1}/\gamma$ for a single pump with the second pump kept constant ($\Omega_{2}/\gamma = 0$). We present here the steady entangled states at different interatomic distances $r_{12}/\lambda_{0} = 0.5$, $r_{12}/\lambda_{0} = 1.0$, and $r_{12}/\lambda_{0} = 1.5$, respectively. The heatmap shows a high concurrence at $r_{12}/\lambda_{0} = 0.5$ which corresponds to the high decay ratio $\gamma_{12}/\gamma$ and minimal dipole-dipole interaction $g_{12}$ which satisfy the aforementioned criteria for qubits entanglement. It can be seen that as the decay ratio $\gamma_{12}/\gamma$ decreases relative to the interatomic distances $r_{12}/\lambda_{0}$, the concurrence decreases. Note that the normalized interatomic distance $r_{12}/\lambda_{0} = 0.5$, $r_{12}/\lambda_{0} = 1.0$, and $r_{12}/\lambda_{0} = 1.5$ correspond to a qubit separation of $r_{12} = 725$ nm, $r_{12} = 1450$ nm, and $r_{12} = 2175$ nm, respectively, which exceeds the resonance energy transfer of quantum emitters in vacuum.

Figure \ref{fig:fig7a} (d) - (f) also shows the concurrence of quantum emitters in the presence of a coherent pump source at three normalized interatomic distances $r_{12}/\lambda_{0} = 0.5$, $r_{12}/\lambda_{0} = 1.0$, $r_{12}/\lambda_{0} = 1.5$. Similar calculations for vacuum reservoir ($r_{12}/\lambda_{0} = 0.5$) (See Appendix for details) are illustrated in the corresponding figures for reference purposes. We implement three types of pumps namely, asymmetric with Rabi frequencies $\Omega_{1} \neq 0, \Omega_{2}=0$, symmetric with $\Omega_{1}=\Omega_{2}$, and antisymmetric pumping with Rabi frequencies $\Omega_{1}=-\Omega_{2}$. As expected, we obtain high steady-state concurrence at the $r_{12}/\lambda_{0} = 0.5$ to $r_{12}/\lambda_{0} = 1.0$, as compared with the vacuum medium. This shows that the ENZ rolled-up waveguide attains high entangled states at the cutoff wavelength and enhance long-range interactions of quantum emitters. From all the aforementioned pumping states, we see a persistent concurrence as a result of the external source to compensate for the depopulation of the entangled states. Figure \ref{fig:fig7a} (d) and (f) (i.e. asymmetric and antisymmetric pumping) show quite identical steady-state concurrence with high entangled states in the proposed ENZ rolled-up waveguide at different interatomic distances $r_{12}/\lambda_{0}$. In addition, the symmetric pumping criteria sustains the concurrence only to normalized time $t\gamma = 15$ at $r_{12}/\lambda_{0} = 0.5$. The high and persistent concurrence for the different pumping criteria (i.e. asymmetric and antisymmetric pumping) illustrates that one can achieve steady entangled states $C(t \rightarrow \infty)$ by compensating the depopulation of the excited states through pumping the qubits with a coherent external source (i.e. monochromatic laser source). 

\section{Conclusion}
We have presented the numerical calculations of the ENZ mode-mediated entanglement between two quantum emitters coupled to a rolled-up ENZ waveguide. The numerical calculations demonstrate that, at the cutoff wavelength, the rolled-up ENZ waveguide can serve as a reservoir to mediate dipole-dipole interactions and long-range entangled states. Moreover, the calculations establish that the rolled-up ENZ waveguide design will enhance resonance energy transfer and transient entanglement of qubits at the cutoff wavelength. We also demonstrate that the transient entanglement of qubits mediated by this novel rolled-up ENZ waveguide could reach its steady-state by using an external pump source. The different pumping systems compensate for the depopulation of the emitter excited states in the rolled-up ENZ waveguide. 

Our design and numerical calculations have established that rolled-up ENZ waveguide can serve as a unique reservoir for quantum entanglement. As a proof of concept, we fabricated the rolled-up ENZ waveguide using a self-rolling mechanism. The practical realization of the proposed design is experimentally novel and relevant to pursue in our subsequent works. We also envision that this rolled-up ENZ waveguide could overcome the practical challenges of quantum emitters integration. This will open a new avenue to explore entanglement in ENZ mediums practically which is promising for quantum teleportation, computing, and communication.
\newpage
\medskip
\textbf{Conflicts of interest}\par
There are no conflicts to declare.

\medskip
\textbf{Acknowledgements} \par 
We acknowledge the financial support of the European Research Council (Starting Grant project aQUARiUM; Agreement No. 802986), Academy of Finland Flagship Programme, (PREIN), (320165).

\medskip
\textbf{Available supporting information}\par
The version of this code is used to generate the transient entanglement calculations presented in the paper. The codes are provided in this \href{https://github.com/issahi62/Rolled-up-ENZ-Waveguide}{link (i.e. https://github.com/issahi62/Rolled-up-ENZ-Waveguide)} and users of this code are kindly requested to cite its use in their work.

\medskip
\bibliographystyle{MSP}
\bibliography{rsc}

\newpage
\appendix
\renewcommand{\thefigure}{A\arabic{figure}}
\setcounter{figure}{0}
\renewcommand{\theequation}{A.\arabic{equation}}
\setcounter{equation}{0}
\section*{A. Unbounded vacuum dipole-dipole coupling and decay rate}

\begin{figure}[!htb]
\includegraphics[width=\linewidth]{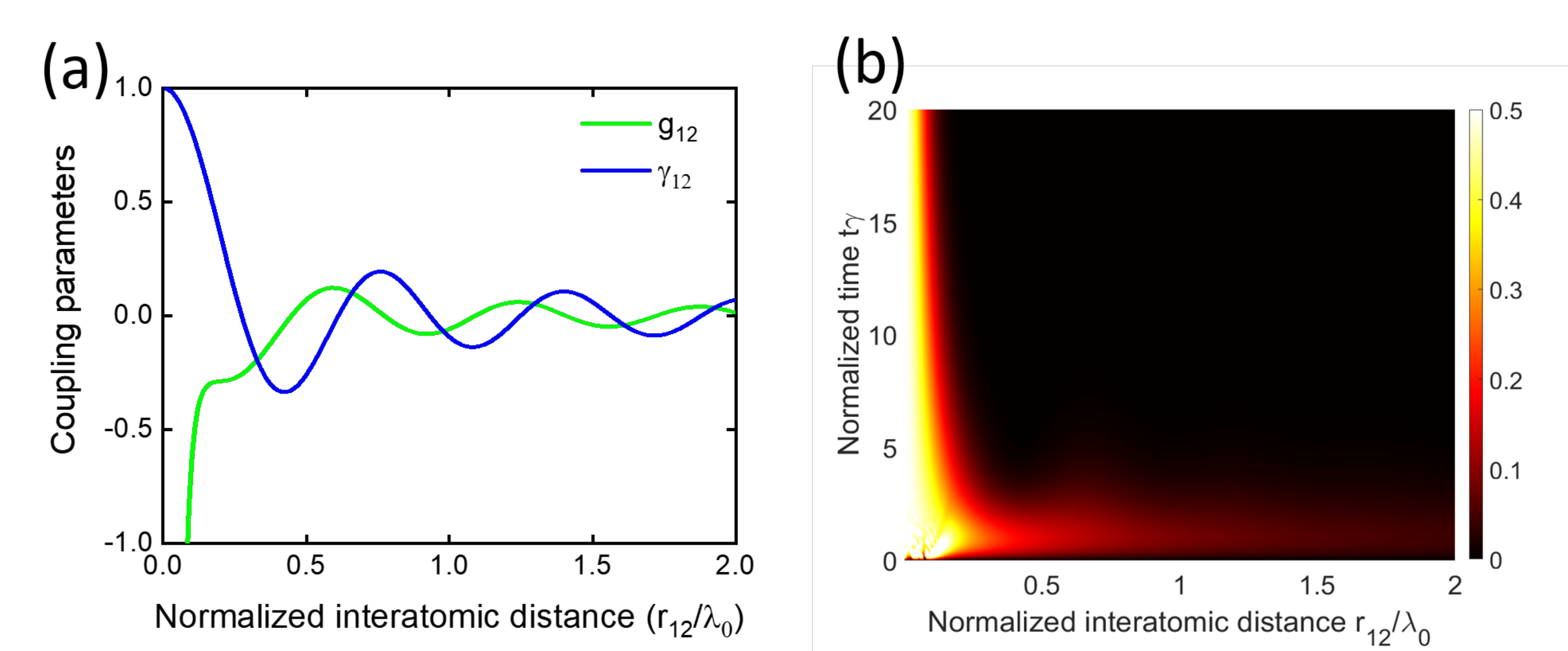}
\caption{\label{fig:fig2} (a) The decay rate and dipole-dipole interaction of a qubit in a vacuum medium. $\gamma_{12}$ is the decay rate as function of interatomic distance ($r$) and $g_{12}$ is the atomic interaction in vacuum. (b) The transient concurrence heatmap of qubits in a vacuum medium.}
\end{figure}

As a comparative study, we numerically calculate the dipole-dipole coupling and decay rate of an unbounded vacuum medium to determine the concurrence based on the aforementioned theory. Note, quantum emitters interactions by a coherent exchange of optical fields are coined as dipole-dipole interactions. The latter interactions are dependent on the orientation of atomic dipole moments as well as the qubits separation distance \cite{Ficek2014}. The emission response $\gamma_{12}$ of a qubit in the vicinity of their environment and their corresponding interactions $g_{12}$ \cite{Fleury2013} shown in Fig. \ref{fig:fig2} can be expressed analytically as, 
\begin{equation}
\gamma_{12}=\frac{3}{2} \gamma_{0}\left(\frac{\sin \left(k_{B} r_{12}\right)}{k_{B} r_{12}}+\frac{\cos \left(k_{B} r_{12}\right)}{\left(k_{B} r_{12}\right)^{2}}-\frac{\sin \left(k_{B} r_{12}\right)}{\left(k_{B} r_{12}\right)^{3}}\right),
\label{eqn15}
\end{equation}
where $k_{B}$ is the wavenumber, $r_{12}$ is the interatomic distance and $\gamma_{0}$ represents the free space decay rate. The dipole-dipole interactions $g_{12}$ in the vacuum medium can also be expressed as
\begin{equation}
g_{12}=\frac{3}{4} \gamma_{0}\left(\frac{\cos \left(k_{B} r_{12}\right)}{k_{B} r_{12}}-\frac{\sin \left(k_{B} r_{12}\right)}{\left(k_{B} r_{12}\right)^{2}}-\frac{\cos \left(k_{B} r_{12}\right)}{\left(k_{B} r_{12}\right)^{3}}\right)
\label{eqn16}
\end{equation}

\begin{figure}[!htb]
 \includegraphics[width=\linewidth]{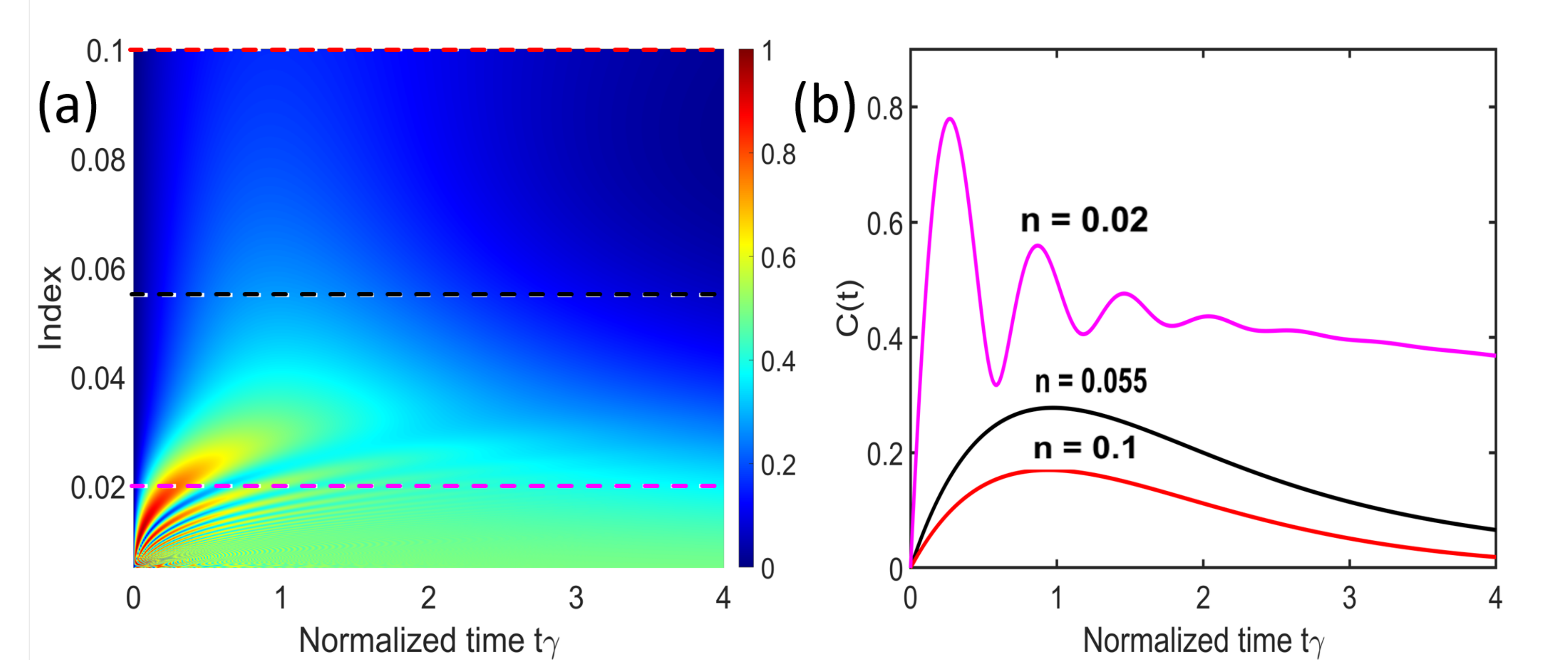}
\caption{\label{fig:fig2bb} (a) The concurrence heatmap as a function of refractive indices (n) and normalized time $t\gamma$. (b) The superimposed dashed lines concurrence plot as shown on the concurrence heatmap.}
\end{figure}

Figure \ref{fig:fig2} (a) illustrates the coupling parameters of quantum emitters in a homogeneous (i.e. air) medium. The decay rate $\gamma_{12}$ and dipole-dipole interactions $g_{12}$ depict the short-range nature of quantum emitters in a vacuum medium. We observed that the free space cooperative emission $\gamma_{12}$ of the quantum emitters is 0.5 as a function of normalized inter-atomic distance $r_{12}/\lambda_{0}$. To comprehend the relevance of using ENZ reservoirs to enhance qubits entanglement, we display the concurrence heatmap shown in Fig. \ref{fig:fig2} (b) using the Wootters formalism \cite{Wootters1998}. The heatmap illustrates weakly entangled states of the qubits as a function of $r_{12}/\lambda_{0}$ and normalized time $t\gamma$. As expected, the short-range decay rate and dipole interactions qubits in a homogeneous medium affect their corresponding entanglement. 
However, incorporating a reservoir with a low index may lead to high concurrence and thereby the inception of ENZ metamaterials and plasmonic channels. Figure \ref{fig:fig2bb} (a) shows the concurrence plot as a function of different refractive indexes (n). The concurrence heatmap and the corresponding crossed section plot illustrated in Fig. \ref{fig:fig2bb} (b) show a high concurrence with a media of minimal refractive index. This provides key insight on utilizing rolled-up ENZ waveguides which possess minimal index at the cutoff wavelength to enhance entanglement between quantum emitters placed at a further distance. 

Also, with respect to the semi-classical phenomena of quantum emitters in the case of unbounded vacuum and waveguide channel evolutions \cite{Sokhoyan2013, Fleury2013}, the decay rate can be expressed as the sum of the contribution of the asymmetrical and symmetrical channels. Analytical formulation of the symmetric $n_{s}$ and antisymmetric $n_{a}$ channels can be expressed as, 
\begin{equation}
n_{s}(t)=\left(1+\frac{\gamma_{12}}{\gamma}\right)\left[1-\left(1-\frac{\gamma_{12}}{\gamma}\right)^{-1} e^{-2 \gamma t}\left(e^{(\gamma-\gamma_{12}) t}-\frac{\gamma_{12}}{\gamma}\right)\right]
\label{eqn17}
\end{equation}
\begin{equation}
n_{a}(t)=\left(1-\frac{\gamma_{12}}{\gamma}\right)\left[1-\left(1+\frac{\gamma_{12}}{\gamma}\right)^{-1} e^{-2 \gamma t}\left(e^{(\gamma+\gamma_{12}) t}+\frac{\gamma_{12}}{\gamma}\right)\right],
\label{eqn18}
\end{equation}
where $\gamma$ is known as the individual decay rate and the collective damping constant is expressed as $\gamma_{12}$. These practical implementations of quantum states with ENZ channels help in developing ways to control and manipulate quantum systems for information processing and energy harvesting with high fidelity. 

\renewcommand{\thefigure}{B\arabic{figure}}
\setcounter{figure}{0}
\renewcommand{\theequation}{B.\arabic{equation}}
\setcounter{equation}{0}
\section*{B. Modelling}
The mode profile, dyadic Green's function, and the entanglement calculations of the proposed design were implemented using 3D Finite-Difference Time-Domain (FDTD) Lumerical software. Based on the attained decay rate and the dipole-dipole interactions, the transient entanglement of the emitter embedded in the proposed design and density matrix is attained using an in-house code developed in MATLAB.  Further analysis was implemented to calculate the steady-state entanglement at the cutoff wavelength of the structure. The simulation was set up by building up 24 alternating layers of SiO\textsubscript{2} and Au with a hollow air core using Lumerical FDTD script commands. The waveguide comprises different alternating metal and dielectric layers with a thickness of 10nm and 5nm respectively.
The complex refractive index of the Au and SiO\textsubscript{2} used in the simulation is from the material data of Johnson and Christy \cite{Johnson1972}, and Palik \cite{palik1997handbook}, respectively. 
The mode profiles and the cutoff frequency of the rolled-up waveguide were determined for different core diameters from 600 nm to 800 nm with a step size of 100 nm. 
The cylindrical hollow multilayered waveguide was built to determine the dyadic Green's response of a quantum emitter embedded within the structure. A quantum emitter is placed in the center of the rolled-up ENZ waveguide oriented in the y-direction. The emitter couples with the ENZ mode which enhances the energy transfer rate from one qubit to the other. A power monitor is placed along the polarization of the dipole source to record the Green's function relative to the coupling between the quantum emitter and the rolled-up ENZ waveguide region. From Eqn. (\ref{eqn2}), the dyadic Green's function based on the above orientation can be expressed as
\begin{equation}
G_{yy} = \frac{\epsilon_{0}\epsilon_{r}c^{2}}{\mu\omega^{2}}E_{y}.
\label{eqn4}
\end{equation}
Note, the decay rate enhancement and the dipole-dipole interaction of the qubits embedded in the rolled-up ENZ waveguides are obtained by the imaginary and the real part of the dyadic Green's functions. 

The local density of states expressed as 
\begin{equation}
\rho_{z}\left(r_{0}, \omega\right)=\frac{6 \omega}{\pi c^{2}}\left[\operatorname{Im}\left\{G_{y y}\left(r_{0}, r_{0}, \omega\right)\right\}\right],
\label{eqn5}
\end{equation}
can also be used to determine the spontaneous decay rate for qubit systems at position $r_{0}$ with their respective transition frequencies.
However, at different orientations of the dipole, the total local density of states can be expressed as the trace of the dyadic Green's function which can also be expressed as 
\begin{equation}
\rho_{z}\left(r_{0}, \omega\right)=\frac{2 \omega}{\pi c^{2}} \operatorname{Im}\left\{\operatorname{Tr}\left[\bar{G}\left(r_{0}, r_{0} ; \omega\right)\right]\right\}.
\label{eqn6}
\end{equation}
The simulation region is set to a 3D layout with a background index set to 1. The mesh refinement was set to conformal variant 0 with a minimum mesh set of 0.25 nm. An additional mesh was used to increase the step size of the simulation and enhance resolution. Different boundary conditions were utilized based on the simulation setup (i.e. Lumerical FDTD, finite difference eigenmode (FDE), and finite element eigenmode (FEEM) solvers) to determine the corresponding results. The number of simulation boundary layers and simulation time is increased to ensure that there is sufficient time for the radiated field to decay completely. 

\renewcommand{\thefigure}{C\arabic{figure}}
\setcounter{figure}{0}
\renewcommand{\theequation}{C.\arabic{equation}}
\setcounter{equation}{0}
\section*{C. Quantum master equation for single qubits}

\begin{figure*}[!htb]
\includegraphics[width=\linewidth]{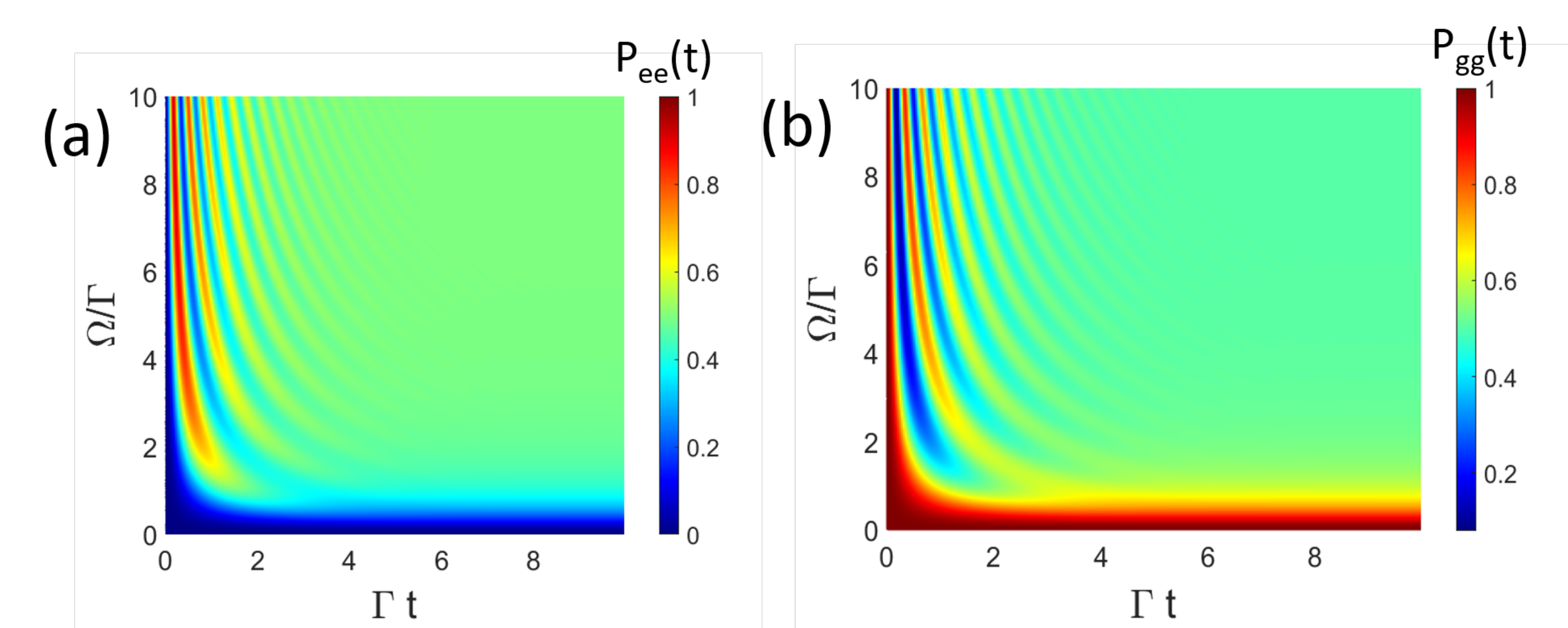}
\caption{\label{fig:figS3}Density state in the presence of a pump laser with variations in Rabi frequency. (a) - (b) The excited and ground state population dynamics as a function normalized time and normalized Rabi frequency.}
\end{figure*}

To determine the population dynamics of a single qubit as a basis to determine the system of equations for two qubits, we illustrated the quantum master formalism for a single qubit \cite{scully_zubairy_1997, Gangaraj2015} which is expressed as
\begin{equation}
\frac{\partial \tilde{\rho}_{S}(t)}{\partial t}=-\frac{i}{\hbar}\left[\tilde{V}^{A F}, \tilde{\rho}_{S}(t)\right]+\mathcal{L} \tilde{\rho}_{S}(t)
\end{equation}
where
\begin{equation}
\mathcal{L} \tilde{\rho}_{S}(t)=\frac{\Gamma\left(\omega_{0}\right)}{2}\left(2 \hat{\sigma} \tilde{\rho}_{s}(t) \hat{\sigma}-\hat{\sigma}^{\dagger} \hat{\sigma} \tilde{\rho}_{s}(t)-\tilde{\rho}_{s}(t) \hat{\sigma}^{\dagger} \hat{\sigma}\right)
\end{equation}
Interaction operator $V_{A F}$ is formulated as 
\begin{equation}
V_{A F}=-\hat{\mathbf{p}} \cdot \mathbf{E}=-\left(\mathbf{p}_{eg} \sigma^{\dagger}+\mathbf{p}_{ge} \sigma\right) \cdot\left(\mathbf{E}_{0} e^{-i \omega t}+\mathbf{E}_{0}^{*} e^{i \omega t}\right)
\end{equation}
which takes the form
\begin{equation}
\begin{aligned}
\tilde{V}_{A F} &=-\left(\mathbf{p}_{eg} \sigma^{\dagger} e^{i \omega_{0} t}+\mathbf{p}_{g e} \sigma e^{-i \omega_{0} t}\right) \cdot\left(\mathbf{E}_{0} e^{-i \omega t}+\mathbf{E}_{0}^{*} e^{i \omega t}\right) \\
& \approx-\mathbf{p}_{eg} \cdot \mathbf{E}_{0} e^{i\left(\omega_{0}-\omega\right) t} \sigma^{\dagger}+\mathbf{p}_{ge} \cdot \mathbf{E}_{0}^{*} e^{i\left(\omega-\omega_{0}\right) t} \sigma \\
&=-\hbar\left(\Omega e^{-i \Delta t} \sigma^{\dagger}+\Omega^{*} e^{i \Delta t} \sigma\right)
\end{aligned}
\end{equation}
where $\Omega=\mathbf{p}_{e g} \cdot \mathbf{E}_{0} / \hbar$ is known as the Rabi frequency and $\Delta=\omega-\omega_{0}$ is the detuning parameter. The matrix elements of the interaction Hamiltonian in the basis $|e\rangle,|g\rangle$, are expressed as system of equations which takes the form
\begin{equation}
\begin{array}{l}
\frac{\partial \rho_{e e}(t)}{\partial t}=-\Gamma \rho_{e e}(t)+i\left(\Omega e^{-i \Delta t} \rho_{g e}(t)-\Omega^{*} e^{i \Delta t} \rho_{e g}(t)\right) \\
\frac{\partial \rho_{eg}(t)}{\partial t}=-\gamma \rho_{e g}(t)+i \Omega e^{-i \Delta t}\left(\rho_{g g}(t)-\rho_{e e}(t)\right) \\
\frac{\partial \rho_{g e}(t)}{\partial t}=-\gamma \rho_{g e}(t)-i \Omega^{*} e^{i \Delta t}\left(\rho_{g g}(t)-\rho_{e c}(t)\right) \\
\frac{\partial \rho_{g g}(t)}{\partial t}=\Gamma \rho_{e e}(t)+i\left(\Omega^{*} e^{i \Delta t} \rho_{e g}(t)-\Omega e^{-i \Delta t} \rho_{ge}(t)\right)
\end{array}
\end{equation}
Assuming a zero detuning parameter: $\Delta=0$, the system of equations can be expressed as
\begin{equation}
\begin{array}{l}
\frac{\partial \rho_{e e}(t)}{\partial t}=-\Gamma \rho_{e e}(t)+i\left(\Omega \rho_{ge}(t)-\Omega^{*} \rho_{e g}(t)\right) \\
\frac{\partial \rho_{e g}(t)}{\partial t}=-\gamma \rho_{e g}(t)+i \Omega e^{-i \Delta t}\left(\rho_{g g}(t)-\rho_{e e}(t)\right) \\
\frac{\partial \rho_{ge}(t)}{\partial t}=-\gamma \rho_{ge}(t)-i \Omega^{*} e^{i \Delta t}\left(\rho_{g g}(t)-\rho_{e e}(t)\right) \\
\frac{\partial \rho_{g g}(t)}{\partial t}=\Gamma \rho_{e e}(t)+i\left(\Omega^{*} \rho_{e g}(t)-\Omega \rho_{g e}(t)\right)
\end{array}
\end{equation}
Also, assuming that $\Omega=\Omega^{*}$ and taking into account that $\rho_{e g}=\rho_{e g}^{\prime}+i \rho_{eg}^{\prime \prime} .$ The system of equations can be expressed as
\begin{equation}
\begin{array}{l}
\frac{\partial \rho_{e e}(t)}{\partial t}=-\Gamma \rho_{e e}(t)+2 \Omega \rho_{e g}^{\prime \prime}(t) \\
\frac{\partial \rho_{e g}^{\prime \prime}(t)}{\partial t}=-\Gamma \rho_{e g}^{\prime \prime}(t)+\Omega\left(1-2 \rho_{e e}(t)\right)
\end{array}
\end{equation}
with the assumption that $\Gamma=\gamma$, and $\rho_{g g}=1-\rho_{ee}$. 
The solution of the system of equations can then be expressed as
\begin{equation}
\rho_{e e}(t)=C_{e}(t) e^{-\Gamma t}, \quad \rho_{e g}^{\prime \prime}(t)=C_{e g}^{\prime \prime}(t) e^{-\Gamma t}
\end{equation}
where
\begin{equation}
\begin{aligned}
\frac{\partial C_{e}(t)}{\partial t} &=2 \Omega C_{e g}^{\prime \prime}(t) 
\frac{\partial C_{e g}^{\prime \prime}(t)}{\partial t} &=\Omega e^{\Gamma t}\left(1-2 C_{e}(t) e^{-\Gamma t}\right)
\end{aligned}
\end{equation}
Thus
\begin{equation}
\frac{\partial^{2} C_{e}(t)}{\partial t^{2}}=2 \Omega^{2} e^{\Gamma t}-4 \Omega^{2} C_{e}(t)
\end{equation}
The general solution is a sum of a homogeneous and a particular solution which is expressed as 
\begin{equation}
C_{e}(t)=C_{0} \cos (2 \Omega t)+C_{1} \sin (2 \Omega t)+\frac{2 \Omega^{2} e^{\Gamma t}}{\Gamma^{2}+4 \Omega^{2}}
\end{equation}
To solve for $C_{0}$ and $C_{1}$, we assume that our system was initially in the ground state, i.e. $\rho_{ee}(0)=0$. Then
\begin{equation}
0=C_{0}+\frac{2 \Omega^{2}}{\Gamma^{2}+4 \Omega^{2}} \quad \Rightarrow \quad C_{0}=-\frac{2 \Omega^{2}}{\Gamma^{2}+4 \Omega^{2}}
\end{equation}
In order to find $C_{1}$ we need to use initial condition $\rho_{e g}^{\prime \prime}(0)=0$. Then
\begin{equation}
\begin{aligned}
C_{e}(t) &=-\frac{2 \Omega^{2}}{\Gamma^{2}+4 \Omega^{2}} \cos (2 \Omega t)+C_{1} \sin (2 \Omega t)+\frac{2 \Omega^{2} e^{\Gamma t}}{\Gamma^{2}+4 \Omega^{2}} 
C_{e g}^{\prime \prime}(0) &=\left.\frac{1}{2 \Omega} \frac{\partial C_{e}(t)}{\partial t}\right|_{t=0}= C_{1} \cos (2 \Omega t)+\left.\frac{\Omega \Gamma e^{\Gamma t}}{\Gamma^{2}+4 \Omega^{2}}\right|_{t=0}\nonumber \\
 &=C_{1}+\frac{\Omega \Gamma}{\Gamma^{2}+4 \Omega^{2}}=0
\end{aligned}
\end{equation}
which can be simplified as
\begin{equation}
C_{1}=-\frac{\Omega \Gamma}{\Gamma^{2}+4 \Omega^{2}}
\end{equation}
The final density matrix can be expressed as
\begin{equation}
\rho_{e e}(t)=\frac{-\Omega(2 \Omega \cos (2 \Omega t)+\Gamma \sin (2 \Omega t)) e^{-\Gamma t}+2 \Omega^{2}}{\Gamma^{2}+4 \Omega^{2}}
\end{equation}
which can be normalized to dimensionless units $\tilde{t}=\Gamma t, \Omega=\alpha \Gamma$. This takes the form
\begin{equation}
\begin{aligned}
\rho_{e e}(t)=\frac{-\alpha(2 \alpha \cos (2 \alpha \Gamma t)+\sin (2 \alpha \Gamma t)) e^{-\tilde{t}}+2 \alpha^{2}}{1+4 \alpha^{2}}\nonumber 
=\frac{2 \alpha^{2}}{1+4 \alpha^{2}}-\frac{\alpha e^{-\bar{t}}}{1+4 \alpha^{2}}(2 \alpha \cos (2 \alpha \tilde{t})+\sin (2 \alpha \tilde{t}))
\end{aligned}
\end{equation}
and the corresponding heatmaps for both the ground and excited population dynamics is shown in Fig.\ref{fig:figS3}.

\renewcommand{\thefigure}{D\arabic{figure}}
\setcounter{figure}{0}
\renewcommand{\theequation}{D.\arabic{equation}}
\setcounter{equation}{0}
\section*{D. Fabrication}
The cleaned silicon (Si) substrate is coated with a monolayer of Hexamethyldisilazane (HMDS) at 125$^{\circ}$ Celsius (C)  to improve the adhesion of the photoresist. Then, we spin coat image reversal resist AZ5214E at 3000 rounds per minute for 40 seconds (sec) and soft bake at 100 $^{\circ}$C. The coated samples are exposed under ultraviolet light (UV) using a  Suss MA6 mask aligner for 4 s. The exposed samples are baked at 115 $^{\circ}$C for 2 minutes (min). For the image reversal process, the samples are flood exposed to UV light for 30 seconds without a mask. The rectangular pattern (25X25 $\mu$m$^{2}$) appears once the exposed samples were developed for 60 sec using MIF 726 developer and then rinsed three times in di-ionized (DI) water. The developed samples are coated with 40 nm of Ge which serves as a sacrificial layer. The unwanted material is removed using the lift-off process, by leaving the samples in S1165 remover at 80 $^{\circ}$ C and sonicated for 2 min. 

\begin{figure*}[!htb]
\includegraphics[width=\linewidth]{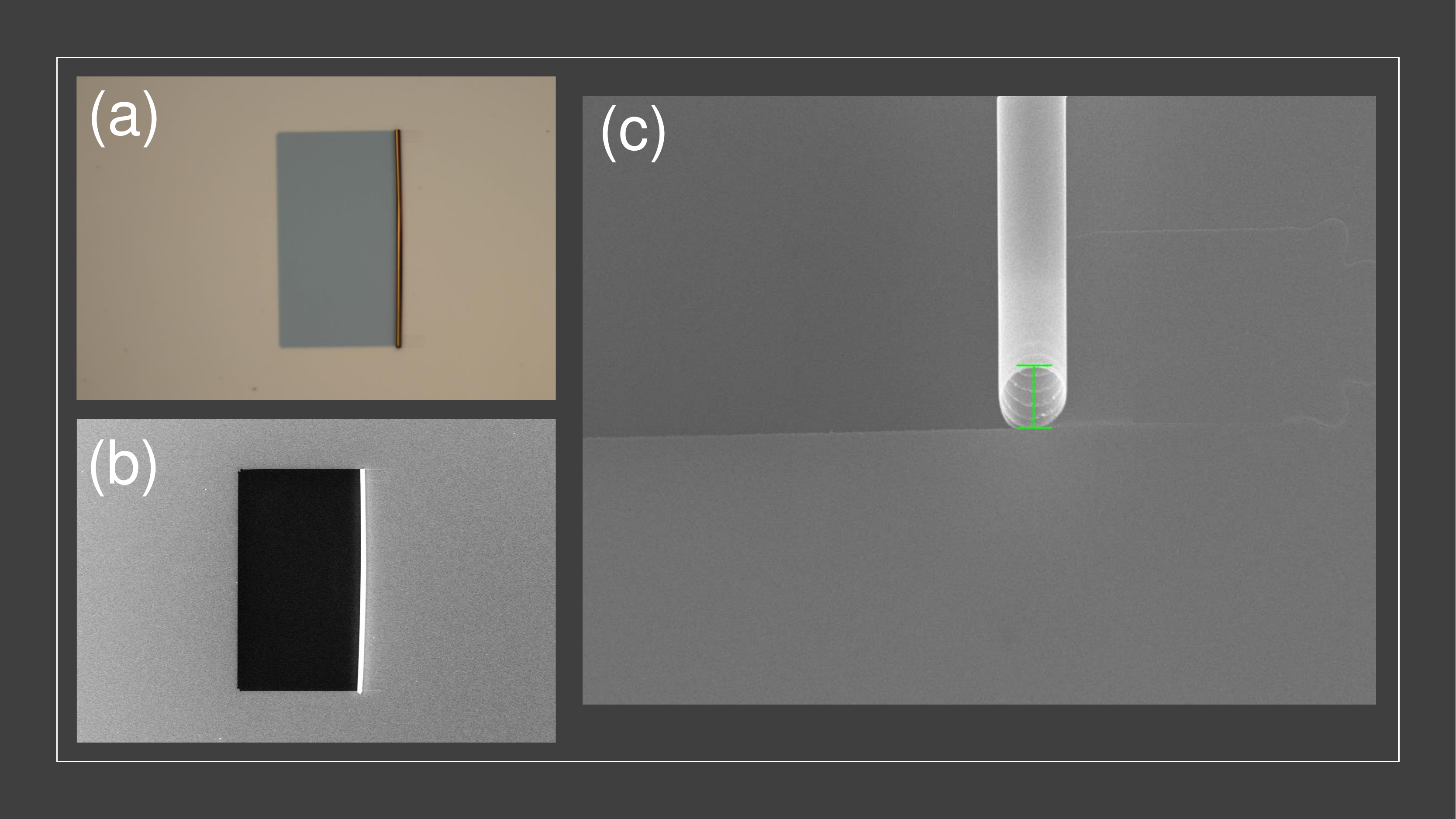}
\caption{\label{fig:fig12} (a) Microscope and scanning electron microscope image (b) of the full rolled-up ENZ waveguide from the top and (c) represents the cross-section with a diameter of 700 nm.}
\end{figure*}

Once rectangular structures of 40 nm of Ge are achieved the sample is again coated with HMDS and positive resist AZ3012E with the same speed and time. The coated samples are soft baked at 90 $^{\circ}$C for 90 sec and exposed to a second mask to create a window for etching the Ge layer. The exposed samples are hard-baked at 110 $^{\circ}$C for 60 sec. This time samples are developed for 1 minute in MIF 726 developer and rinsed under the DI water. The developed samples are coated with 5 nm of SiO\textsubscript{2}, 1 nm of titanium (Ti) as an adhesive layer, and 10 nm of Au. The difference in the deposition rate and materials density during the deposition process inherits different strains. The SiO\textsubscript{2} undergoes compressive stress, while Au with a high deposition rate will have tensile stress. The two opposite stress conditions when released by etching the bottom Ge layer for 120 min in 35$\%$ H\textsubscript{2}O\textsubscript{2}, will start rolling the SiO\textsubscript{2} and Au in the upward direction. The thicknesses of the SiO\textsubscript{2}/Au and the use of Ge as etch layer tightly roll the tubes to achieve a small diameter of 700 nm, reported for the first time. Microscope and scanning electron microscope image of the full rolled-up ENZ waveguide is shown in Fig. \ref{fig:fig12}.

\newpage

\begin{figure}
\textbf{Table of Contents}\\
\medskip
  \includegraphics{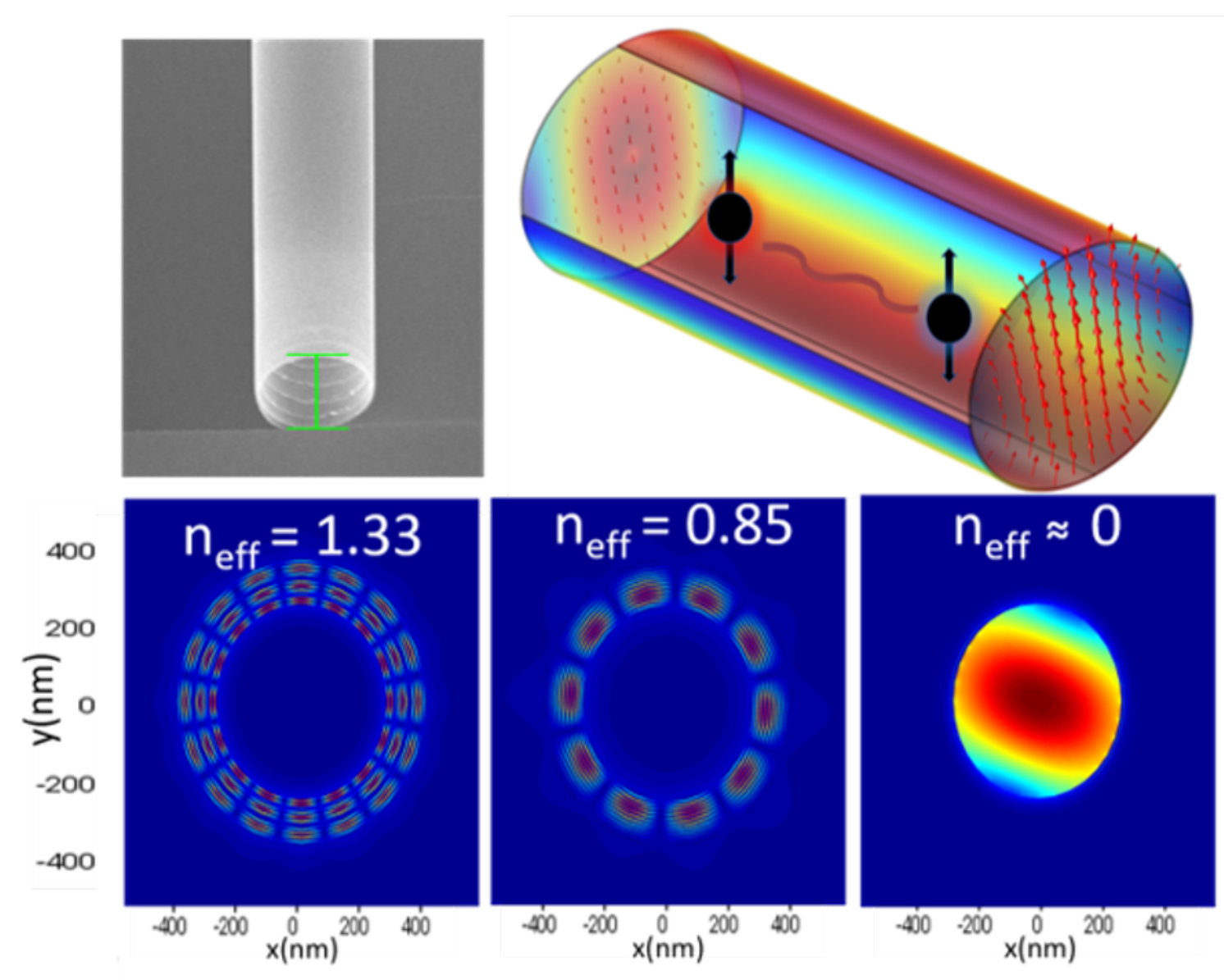}
  \medskip
  \caption*{ToC Entry}
\end{figure}

\end{document}